\documentclass[pra,onecolumn,floatfix,a4paper,superscriptaddress]{revtex4}
\usepackage{bm,color,graphicx,amsmath,txfonts}
\usepackage{here}
\usepackage{array}
\usepackage{graphicx}
\usepackage[colorlinks, citecolor=blue,linkcolor=blue]{hyperref}
\usepackage{braket}
\usepackage{dsfont}
\usepackage[left=	2cm,top=2cm,right=1.2cm,bottom=2cm]{geometry}
\usepackage{tikz}

\begin{document}

\makeatletter
\renewcommand{\@biblabel}[1]{\makebox[2em][l]{\textsuperscript{\textcolor{black}{\fontsize{10}{12}\selectfont[#1]}}}}
\makeatother

\let\oldbibliography\thebibliography
\renewcommand{\thebibliography}[1]{%
  \addcontentsline{toc}{section}{\refname}%
  \oldbibliography{#1}%
  \setlength\itemsep{0pt}%
}

\title{Dephasing Effects on the Dynamical Evolution of Quantum Correlations and Coherence in Neutrino Oscillations}

\author{Omar Bachain}
\address{LPHE-Modeling and Simulation, Faculty of Sciences, Mohammed V University in Rabat, Rabat, Morocco}

\author{Elhabib \surname{Jaloum} }
\address{LPTHE-Department of Physics, Faculty of Sciences, Ibnou Zohr University, Agadir 80000, Morocco}

\author{Mohamed \surname{Amazioug} }
\email{m.amazioug@uiz.ac.ma}
\address{LPTHE-Department of Physics, Faculty of Sciences, Ibnou Zohr University, Agadir 80000, Morocco}

\author{Nazek Alessa}
\address{Department of Mathematical Sciences, College of Science, Princess Nourah bint Abdulrahman University, P.O. Box 84428, Riyadh 11671, Saudi Arabia}

\author{Wedad R. Alharbi}
\address{Physics Department, College of Science, University of Jeddah, Jeddah, 23890, Saudi Arabia}

\author{Rachid Ahl Laamara}
\address{LPHE-Modeling and Simulation, Faculty of Sciences, Mohammed V University in Rabat, Rabat, Morocco}
\address{Centre of Physics and Mathematics, CPM, Faculty of Sciences, Mohammed V University in Rabat, Rabat, Morocco}

\author{Abdel-Haleem Abdel-Aty}
\affiliation{Department of Physics, College of Sciences, University of Bisha, Bisha 61922, Saudi Arabia}

\date{\today}

\begin{abstract}

Neutrino oscillations confirm the presence of mode entanglement, as each flavor eigenstate is composed of a coherent superposition of distinct mass eigenstates. In this work, we investigate the dynamics of quantum resources in neutrino oscillation systems by analyzing quantum steering, logarithmic negativity, and quantum coherence within a two-flavor framework. Treating neutrino oscillations as an effective two-level quantum system, we study the influence of environmental decoherence on these nonclassical features by modeling the system as an open quantum system. Three representative noise channels are considered, namely amplitude damping (AD), phase flip (PF), and phase damping (PD), allowing us to capture both dissipative and dephasing mechanisms. We examine the evolution of quantum resources in both Markovian and non-Markovian regimes, highlighting the role of memory effects in the system-environment interaction. The results reveal a clear hierarchy in the robustness of quantum resources under decoherence. Steering is the most sensitive correlation in the hierarchy under decoherence effects. while logarithmic negativity exhibits intermediate robustness. Quantum coherence displays the highest resilience, persisting over a wider range of parameters. In the PF and PD channels, logarithmic negativity and coherence are shown to exhibit identical dynamical behavior, reflecting their common dependence on phase-related noise. In contrast, the non-Markovian regime leads to delayed decoherence and partial revivals of entanglement and coherence due to information backflow, whereas quantum steering remains strongly suppressed. These findings provide a comparison of different quantum resources in neutrino oscillation systems and offer new insights into the interplay between decoherence mechanisms and quantum correlations.

\end{abstract}
\maketitle

\section{Introduction}    \label{sec:1}

Quantum nonlocality was first introduced by Einstein, Podolsky, and Rosen in 1935, when they introduced a thought experiment to scrutinize the completeness of the quantum formalism \cite{EPR1935}. This argument, later termed the EPR paradox, revealed the existence of strong correlations between spatially separated quantum systems that defy classical local explanations, suggesting that measurements performed on one subsystem could instantaneously influence the state of another, regardless of the distance. To clarify these nonclassical features, Schr\"odinger introduced the concepts of entanglement and steering, providing a rigorous framework for characterizing nonlocal correlations \cite{Schrodinger1935}. Subsequently, Bell demonstrated that no theory based on local hidden variables can reproduce all the predictions of quantum mechanics, thereby establishing the fundamentally nonlocal nature of the theory \cite{Bell1964}. The experimental violation of Bell inequalities was later confirmed by Aspect and collaborators through landmark experiments employing entangled photon pairs, offering definitive empirical support for quantum mechanics \cite{Aspect1981,Aspect1982}. As a purely quantum phenomenon with no classical counterpart, entanglement generates correlations within multipartite systems and constitutes a fundamental resource for a wide range of quantum information processing tasks \cite{Horodecki2009}.\\%

Beyond Bell nonlocality, distinct forms of quantum correlations have been identified and organized within a hierarchical framework, among which quantum steering plays a central role. Formally introduced by Wiseman \textit{et al.}, quantum steering characterizes the ability of one party to nonlocally influence the state of another through local measurements, occupying an intermediate position between entanglement and Bell nonlocality \cite{Wiseman2007}. This inherent asymmetry makes steering particularly relevant for one-sided device-independent quantum information protocols, such as secure quantum communication and quantum key distribution \cite{Cavalcanti2015}. Entanglement, as the fundamental resource underlying steering, is commonly quantified by the logarithmic negativity, which provides an operationally meaningful and computationally efficient measure for mixed states and open quantum systems \cite{Vidal2002,Plenio2005}. In parallel, quantum coherence has been recognized as an essential manifestation of the superposition principle and a resource in its own right, rigorously formalized within the resource-theoretic approach \cite{Baumgratz2014}. Quantifiers such as the $l_1$-norm of coherence capture the degree of basis-dependent superposition present in a quantum state. Importantly, coherence and quantum correlations are deeply interconnected; coherence can be converted into entanglement and steering through appropriate interactions, highlighting its foundational role in the emergence of nonclassical correlations \cite{Streltsov2017}. Understanding the interplay and robustness of these quantum resources under realistic noisy environments therefore constitutes a key challenge for both fundamental quantum theory and practical applications\cite{CavazzoniTekluParis2025,HeParis2025,AsjadTekluParis2023}.\\

In realistic physical settings, quantum systems inevitably interact with their surrounding environments, giving rise to decoherence and dissipation that progressively degrade quantum steering, entanglement, and coherence. Such open-system dynamics are commonly described using quantum noise channels, among which the amplitude damping (AD), phase flip (PF), and phase damping (PD) channels constitute paradigmatic models for energy relaxation and phase randomization processes \cite{NielsenChuang,BreuerPetruccione}. While the AD channel accounts for irreversible population loss due to energy exchange with the environment, the PF and PD channels describe distinct forms of phase noise that affect quantum superpositions without modifying populations \cite{Palma1996}. In addition, the nature of the system–environment interaction crucially determines the temporal behavior of quantum resources. In the Markovian regime, memoryless dynamics typically lead to a monotonic decay of steering, entanglement, and coherence; conversely, in the non-Markovian regime, environmental memory effects can induce information backflow, resulting in delayed decoherence and partial revivals of nonclassical features \cite{Breuer2009,Rivas2014,deVega2017}. A systematic comparison of these noise channels across both dynamical regimes is therefore essential for assessing the robustness and controllability of quantum steering, entanglement, and coherence in realistic quantum systems.\\%
 
In addition to their relevance in quantum information theory, nonclassical correlations arise in fundamental particle physics, particularly in the phenomenon of neutrino oscillations \cite{Pontecorvo1957,Pontecorvo1968,Ettefaghi2020,DixitAlok2021,BlasoneDeSienaMatrella2021,Bittencourt2022,Song2018,Alok2025SpinFlavor,YadavAlok2025}. Neutrinos are produced and detected in flavor eigenstates but propagate as coherent superpositions of mass eigenstates. This leads to oscillatory transitions between different flavors, as described by the mixing formalism introduced by Maki, Nakagawa, and Sakata \cite{MNS1962}. In the present work, we adopt a two-flavor framework, which captures the essential features of neutrino oscillations and allows for a transparent quantum-mechanical description of the system \cite{GiuntiKim}. Within this approach, the neutrino state can be treated as an effective two-level quantum system, providing a natural platform for exploring the interplay between quantum steering, entanglement, and coherence.\\%
 
From this perspective, neutrino oscillations provide a physical realization of quantum superposition and interference, where environmental effects, matter interactions, and stochastic fluctuations can induce decoherence and alter the evolution of quantum resources \cite{Blennow2017}. By embedding the two-flavor neutrino oscillation model within the formalism of open quantum systems, we investigate how different noise channels and dynamical regimes-including both Markovian and non-Markovian processes-influence quantum steering, logarithmic negativity, and quantum coherence \cite{BreuerPetruccione,Breuer2009}. This link between neutrino physics and quantum information theory provides a unified framework for analyzing nonclassical features in realistic physical systems.\\%
 
The remainder of this paper is organized as follows. In Sec.~\ref{sec:2}, we introduce the theoretical model, with particular emphasis on the two-flavor neutrino oscillation framework. Section~\ref{sec:3} is devoted to the definition and quantification of the quantum correlations considered, including quantum steering, logarithmic negativity, and quantum coherence. In Sec.~\ref{sec:4}, we describe the noisy quantum channels analyzed in this work, namely the amplitude damping, phase flip, and phase damping channels. The main results and their physical interpretations are presented and discussed in Sec.~\ref{sec:5}. Section~\ref{sec:6} investigates the dephasing effects and the dynamical behavior of the quantum state under different environmental conditions. In Sec.~\ref{sec:7}, we discuss the experimental feasibility of probing quantum correlations in neutrino systems, with particular emphasis on indirect reconstruction strategies and realistic experimental constraints. Finally, Sec.~\ref{sec:8} is devoted to the conclusions.

\section{Theoretical model}\label{sec:2}
In the two-flavor approximation, neutrino oscillations can be naturally described within
the framework of quantum information by mapping flavor states onto a two-qubit Hilbert
space. A neutrino produced in a definite flavor state evolves as a coherent superposition
of two mass eigenstates, giving rise to single-particle mode entanglement between the
corresponding flavor modes. The origin of this flavor change lies in the existence of a
mixing matrix connecting flavor eigenstates to mass eigenstates. This mixing is formally
described by the unitary Pontecorvo--Maki--Nakagawa--Sakata (PMNS) matrix, such that
\begin{equation}
	|\nu_\alpha^{\,h}(t)\rangle
	=
	\sum_{i=1}^{2}
	U_{\alpha i}\,|\nu_i^{\,h}(t)\rangle.
\end{equation}
The neutrino mass eigenstates 
$\ket{\nu^h_i}$
are modeled as plane waves and therefore obey the Schrödinger equation 
\begin{equation}
	i \frac{\partial}{\partial t} \, |\nu_i^h(t)\rangle = E_i \, |\nu_i^h(t)\rangle.
\end{equation}
After a straightforward calculation, the time evolution of a neutrino flavor state with a well-defined helicity 
$h$ is obtained and is given by \cite{Alok2025Coherent}.
\begin{equation}
	\ket{\nu^h_\alpha(t)} = \sum_{i=1}^{2} U_{\alpha i} e^{-i E_i t} \ket{\nu^h_i(0)},
\end{equation}
where $\alpha,\beta \in \{e,\mu,\tau\}$ denote the flavor indices, $i=1,2$ label the mass eigenstates, and $U_{\alpha i}$ are the elements of the effective $SU(2)$ mixing matrix. In this reduced description, neutrino mixing is fully characterized by a single mixing
angle $\theta$
\begin{equation}
	U_{\alpha i}=\begin{pmatrix}
		\cos\theta & \sin\theta\\
		-\sin\theta &\cos\theta
	\end{pmatrix}.
\end{equation}

In the ultra-relativistic limit, the exact quantum field--theoretic flavor states reduce
to the Pontecorvo states, which represent physically well-defined single-particle states
associated with distinct flavor modes. This allows neutrino oscillations to be interpreted
in terms of quantum correlations between flavor modes.
To establish the qubit description, we adopt the occupation number representation, where
the presence (absence) of a neutrino in a given flavor mode is denoted by the qubit state
$\ket{1}$ ($\ket{0}$), respectively. Accordingly, the flavor states at the production point
$L=0$ are mapped onto the bipartite two-qubit basis as
\begin{equation}
	\ket{\nu_\alpha} = \ket{1}_\alpha \otimes \ket{0}_\beta, \qquad
	\ket{\nu_\beta} = \ket{0}_\alpha \otimes \ket{1}_\beta .
\end{equation}

Under time evolution, an initial flavor state $\ket{\nu_\alpha}$ evolves into a Bell-like
superposition in the composite Hilbert space $\mathcal{H}_2 \otimes \mathcal{H}_2$,
\begin{equation}
	\ket{\nu_\alpha(\theta,L)} =
	\tilde{U}_{\alpha\alpha}(\theta,L)\, \ket{1}_\alpha \otimes \ket{0}_\beta
	+ \tilde{U}_{\alpha\beta}(\theta,L)\, \ket{0}_\alpha \otimes \ket{1}_\beta ,
	\label{eq:twoflavor_state}
\end{equation}
where the complex amplitudes are given by
\begin{align}
	\tilde{U}_{\alpha\alpha}(\theta,L) &=
	\cos^2\theta\, e^{-i E_i L} + \sin^2\theta\, e^{-i E_j L}, \nonumber\\
	\tilde{U}_{\alpha\beta}(\theta,L) &=
	\sin\theta \cos\theta \left( e^{-i E_j L} - e^{-i E_i L} \right).
	\label{eq:twoflavor_coefficients}
\end{align}
These coefficients satisfy the normalization condition
$|\tilde{U}_{\alpha\alpha}|^2 + |\tilde{U}_{\alpha\beta}|^2 = 1$, where
$|\tilde{U}_{\alpha\alpha}|^2$ represents the survival probability of flavor $\alpha$,
and $|\tilde{U}_{\alpha\beta}|^2$ corresponds to the transition probability from $\alpha$
to $\beta$.
From Eq.~\eqref{eq:twoflavor_state}, the two-qubit density matrix describing the
mode-entangled neutrino state in the two-flavor scenario is given by
\begin{equation}
	\varrho
	= \ket{\nu_\alpha(\theta,L)}\bra{\nu_\alpha(\theta,L)},
\end{equation}
which explicitly reads
\begin{equation}
	\varrho=
	\begin{pmatrix}
		0 & 0 & 0 & 0 \\
		0 & \varrho_{22}&
		\varrho_{23} & 0 \\
		0 &\varrho_{32}&
		\varrho_{33}& 0 \\
		0 & 0 & 0 & 0
	\end{pmatrix},
	\label{eq: 6} 
\end{equation}
where 
\begin{align}
	\varrho_{22}=|\tilde{U}_{\alpha\alpha}|^2&=1-\sin^2(2\theta)\sin^2\left(\phi\right) \nonumber\\
	\varrho_{33}=|\tilde{U}_{\alpha\beta}|^2&=\sin^2(2\theta)\sin^2\left(\phi\right) \nonumber\\
	\varrho_{23}=\tilde{U}_{\alpha\alpha}\tilde{U}^*_{\alpha\beta}&=\sin(2\theta)\left[ -\cos(2\theta)\sin^2\left(\phi \right)-i \sin\left(\phi \right)\cos\left(\phi \right)\right] \\
	\varrho_{32}=	\varrho_{23}^*=\tilde{U}^*_{\alpha\alpha}\tilde{U}_{\alpha\beta}=&\sin(2\theta)\left[ -\cos(2\theta)\sin^2\left(\phi \right)+i \sin\left(\phi \right)\cos\left(\phi \right)\right],\nonumber 
\end{align}
where $\phi=\frac{\Delta m_{ij}^2}{4E}L$. We assume that the different mass eigenstates are produced with the same momentum
(the \emph{equal momentum assumption}), $|\mathbf{p}| \simeq p = E$.
Since neutrinos are very light particles, we further assume $m_\nu \ll E$.
Under these assumptions, the energy of the $i$-th mass eigenstate satisfies
\begin{equation}
	E_i^2 = p_i^2 + m_i^2 \simeq E^2 + m_i^2 ,
\end{equation}
which leads to
\begin{equation}
	E_i = \sqrt{E^2 + m_i^2} \simeq E + \frac{m_i^2}{2E} .
\end{equation}
Consequently, the energy difference between two mass eigenstates can be approximated as
\begin{equation}
	E_i - E_j \simeq \frac{m_i^2 - m_j^2}{2E}
	= \frac{\Delta m_{ij}^2}{2E} .
\end{equation}
Moreover, we assume that the produced neutrinos are ultra-relativistic.
Setting $c = 1$, this implies
$L \simeq T $, 
where $T$ is the propagation time and $L$ is the distance traveled by the neutrino,
$\Delta m_{ij}^2$ denotes the corresponding mass-squared difference.
This relation plays a central role in neutrino oscillation physics, as it allows the oscillation
phase to be expressed in terms of the mass-squared differences \( \Delta m_{ij}^2 \), which are
among the key experimentally measurable parameters governing neutrino flavor conversion. In the following, the mass-squared difference \( \Delta m_{ij}^2 \) is hereafter simply denoted by
\( \Delta m^2 \).
The mixing between the two states \( i \) and \( j \), with \( i,j = 1,2,3 \), is characterized by a
mixing angle generically denoted by \( \theta \).
Depending on the neutrino flavors involved, this angle corresponds to one of the standard leptonic
mixing angles appearing in the PMNS matrix, namely \( \theta_{12} \), \( \theta_{23} \), or
\( \theta_{13} \).

\section{QUANTUM RESOURCES MEASURES}\label{sec:3}
\subsection{Quantum Steering}

In this subsection, we quantify the Einstein--Podolsky--Rosen (EPR) steering between the two qubits composing the system. Quantum steering characterizes the ability of one subsystem to nonlocally affect the state of another through local measurements, and it intrinsically exhibits directional asymmetry. We analyze both steering directions: from qubit~1 (Alice, $A$) to qubit~2 (Bob, $B$), and vice versa. The degree of steerability from Alice to Bob is quantified by the normalized steering measure~\cite{CavalcantiSkrzypczyk,AmaziougDaoud2024,JaloumAmazioug2026}
\begin{equation}
	\mathcal{S}_{A \to B} = \max\left(0,\frac{\mathcal{N}_{AB}-2}{4}\right),
	\label{eq:SAtoB}
\end{equation}
where the normalization factor ensures that $\mathcal{S}_{A \to B}=1$ for maximally entangled Bell states. By exchanging the roles of the two subsystems, the steerability from Bob to Alice is given by
\begin{equation}
	\mathcal{S}_{B \to A} = \max\left(0,\frac{\mathcal{N}_{BA}-2}{4}\right).
	\label{eq:SBtoA}
\end{equation}
The steering asymmetry is then defined as
\begin{equation}
	\Delta_{12} = \left| 	\mathcal{S}_{A \to B} -	\mathcal{S}_{B \to A} \right|.
	\label{eq:asymmetry}
\end{equation}

The quantities $\mathcal{N}_{AB}$ and $\mathcal{N}_{BA}$ are constructed from entropic EPR-steering inequalities based on local Pauli measurements $\sigma_x$, $\sigma_y$, and $\sigma_z$. In particular, the steering inequality from Alice to Bob reads
\begin{equation}
	\mathcal{N}_{AB} =
	H(\sigma_x^B|\sigma_x^A)
	+ H(\sigma_y^B|\sigma_y^A)
	+ H(\sigma_z^B|\sigma_z^A)
	\geq 2,
	\label{eq:EPR_AB}
\end{equation}
where $H(B|A)=H(\varrho_{AB})-H(\varrho_A)$ denotes the conditional Shannon entropy.
A violation of this inequality ($\mathcal{N}_{AB}<2$) signals the presence of steering from $A$ to $B$. For the  two-qubit $X$-state density matrix, the quantity $\mathcal{N}_{AB}$ can be evaluated analytically as \cite{AbdRabbou2022,JaloumAmazioug2025} 
\begin{equation}
	\mathcal{N}_{AB} = \frac{1}{2}\sum_{i=1}^{4}
	\left[
	\mathcal{I}^{AB}_{x_i}\log_2 \mathcal{I}^{AB}_{x_i}
	+ \mathcal{I}^{AB}_{y_i}\log_2 \mathcal{I}^{AB}_{y_i}
	+ \mathcal{I}^{AB}_{z_i}\log_2 \mathcal{I}^{AB}_{z_i}
	\right]
	- \sum_{j=x,y,z}\sum_{k}
	\mathcal{I}^{A}_{j_k}\log_2 \mathcal{I}^{A}_{j_k}.
	\label{eq:IAB}
\end{equation}

The joint probability distributions are given by
\begin{align}
	\mathcal{I}^{AB}_{x_{1,2}} = 1 + 2\,\mathrm{Re}(\varrho_{23})&, \quad
	\mathcal{I}^{AB}_{x_{3,4}} = 1 - 2\,\mathrm{Re}(\varrho_{23}), \nonumber\\
	\mathcal{I}^{AB}_{y_{1,2}} = 1 + 2\,\mathrm{Re}(\varrho_{23})&,\quad  
	\mathcal{I}^{AB}_{y_{3,4}} = 1 - 2\,\mathrm{Re}(\rho_{23}),  \\\nonumber
	\mathcal{I}^{AB}_{z_i} = 4\varrho_{ii}.
\end{align}

The marginal probability distributions associated with Alice are
\begin{align}
	\mathcal{I}^{A}_{x_{1,2}} &= 1, \qquad
	\mathcal{I}^{A}_{y_{1,2}} = 1, \nonumber \\
	\mathcal{I}^{A}_{z_{1,2}} &= 1 \pm (\varrho_{22}-\varrho_{33}).
	\label{eq:PA}
\end{align}

Similarly, the steering quantity from Bob to Alice is defined as 
\begin{equation}
	\mathcal{N}_{BA} = \frac{1}{2}\sum_{i=1}^{4}
	\left[
	\mathcal{I}^{AB}_{x_i}\log_2 \mathcal{I}^{AB}_{x_i}
	+ \mathcal{I}^{AB}_{y_i}\log_2 \mathcal{I}^{AB}_{y_i}
	+ \mathcal{I}^{AB}_{z_i}\log_2 \mathcal{I}^{AB}_{z_i}
	\right]
	- \sum_{j=x,y,z}\sum_{k}
	\mathcal{I}^{B}_{j_k}\log_2 \mathcal{I}^{B}_{j_k},
	\label{eq:IBA}
\end{equation}
where Bob’s marginal probabilities are given by
\begin{equation}
	\mathcal{I}^{B}_{z_{1,2}} = 1 \pm (-\varrho_{22}+\varrho_{33}), \quad	\mathcal{I}^{B}_{x_{1,2}}=\mathcal{I}^{B}_{y_{1,2}}=1.
	\label{eq:PB}
\end{equation}

Depending on the values of $	\mathcal{S}_{A \to B}$ and $	\mathcal{S}_{B \to A}$, the system may exhibit one-way steering, two-way steering, or no steering. It is important to emphasize that quantum steering constitutes a stronger form of nonclassical correlation than entanglement: while every steerable state is necessarily entangled, an entangled state is not always steerable. This hierarchy explains the enhanced fragility of quantum steering under thermal decoherence.

\subsection{Logarithmic Negativity}

To quantify entanglement in bipartite mixed states, we employ the logarithmic negativity, $\mathcal{L}_{N}(\varrho)$, as introduced by Vidal and Werner~\cite{Peres1996,VidalWerner2002} and discussed in~\cite{Miranowicz2004}. This measure is based on the sum of the absolute values of the negative eigenvalues of the partially transposed density matrix $\varrho_{T}$, taken with respect to subsystem $Y$~\cite{Plenio2005}. Explicitly, the logarithmic negativity is defined as
\begin{equation}
	\mathcal{L}_{N}(\varrho) = \log_2 \|\varrho_{T}\|_1 
	= \log_2 \left( \sum_i h_i \right),
	\label{eq:LN_def}
\end{equation}
where $h_i$ denote the negative eigenvalues of $\varrho_{T}$, and the trace norm is
\begin{equation}
	\|\varrho_{T}\|_1 = \mathrm{Tr} \sqrt{\varrho_{T} (\varrho_{T})^\dagger}.
	\label{eq:trace_norm}
\end{equation}

For a system of two qubits, the logarithmic negativity can be expressed in terms of the smallest negative eigenvalue $h_{\mathrm{min}}$ of the partially transposed density matrix as
\begin{equation}
\mathcal{L}_{N}(\varrho) = \max \{ 0, -2 h_{\mathrm{min}} \}.
	\label{eq:LN_qubits}
\end{equation}

For the state given in Eq.~(\ref{eq: 6}), the partially transposed density matrix in the computational basis reads
\begin{equation}
	\varrho_{T} =
	\begin{pmatrix}
		0 & 0 & 0 & \varrho_{23} \\
		0 & \varrho_{2,2} & 0 & 0 \\
		0 & 0 & \varrho_{3,3} & 0 \\
		\varrho_{32} & 0 & 0 &0 
	\end{pmatrix},
	\label{eq:rho_PT}
\end{equation}
whose eigenvalues are
\begin{align}
	\vartheta_{1,2} &=  
	\pm 
	\frac{1}{2}\sqrt{  4 |\varrho_{23}|^2 }, \nonumber\\
	\vartheta_{3,4} &= \frac{\varrho_{22} + \varrho_{33}}{2} \pm \frac{1}{2} \sqrt{(\varrho_{22} - \varrho_{33})^2 }.
\end{align}

Finally, the logarithmic negativity is determined from the smallest negative eigenvalue:
\begin{equation}
	\mu_{\mathrm{min}} = \min \{ \vartheta_1, \vartheta_2, \vartheta_3, \vartheta_4 \}.
\end{equation}

\subsection{Quantum Coherence}

Quantum coherence, an essential resource in quantum information processing, can be quantified using the $l_1$-norm of coherence~\cite{Baumgratz2014}. This measure evaluates the degree of coherence of a quantum state by computing the minimum distance between the given state and the set of incoherent states. For a bipartite quantum state 
\begin{equation}
	{\varrho} = \sum_{i,j} \varrho_{ij} \lvert i  \rangle \langle  j \rvert,
\end{equation}

the $l_1$-norm coherence is defined as the sum of the absolute 
\begin{equation}
	\mathcal{C}_{l_1}({\varrho}) = \sum_{i \neq j} \lvert \varrho_{ij} \rvert.
\end{equation}

For the two-qubit system  considered here, the $l_1$-norm coherence can be directly calculated from the density matrix elements as

\begin{equation}
		\mathcal{C}_{l_1}({\varrho}) = \lvert \varrho_{23}\rvert+\lvert \varrho_{32}\rvert.
\end{equation}

\section{NOISY CHANNELS}\label{sec:4}

The interaction between a quantum system and its surrounding environment inevitably induces decoherence, leading to the degradation of quantum coherence and the decay of quantum correlations\cite{Pirandola2008,Damodarakurup2009,AbdRabbou2022}. To describe such open-system dynamics, noisy quantum channels are commonly employed as effective models of environmental interactions. In this work, we consider three standard decoherence channels: amplitude damping (AD), phase flip (PF), and phase damping (PD), each corresponding to a distinct physical noise mechanism \cite{NielsenChuang2010}. The evolution of the bipartite  state $\varrho$ under decoherence is described within the Kraus operator formalism as
\begin{equation}
	{\xi}(\varrho)=\sum_{i,j}{K}_{ij}\,\varrho\,{K}^{\dagger}_{ij},
	\label{eq:Kraus}
\end{equation}
where ${K}_{ij}={K}_{i}\otimes {K}_{j}$ are the Kraus operators acting locally on each subsystem and satisfy the completeness condition
\begin{equation}
	\sum_{i,j}{K}^{\dagger}_{ij}\,{K}_{ij}=\mathbb{I}.
\end{equation}

\subsection{Amplitude Damping Channel}

The amplitude damping (AD) channel describes irreversible energy dissipation processes, such as spontaneous emission, in which an excited qubit decays to its ground state by emitting energy into the environment. The single-qubit Kraus operators are given by
\begin{equation}
	{K}_1=
	\begin{pmatrix}
		1 & 0 \\
		0 & \sqrt{1-\tau}
	\end{pmatrix},
	\qquad
{K}_2=
	\begin{pmatrix}
		0 & \sqrt{\tau} \\
		0 & 0
	\end{pmatrix},
	\label{eq:ADKraus}
\end{equation}
where the decoherence parameter is defined as $\tau=1-e^{-vt}$, with $v$ denoting the decay rate. Applying Eq.~(\ref{eq:Kraus}) to the initial X-type density matrix, the evolved state under the AD channel takes the form
\begin{equation}
	{\varrho}^{AD}=
	\begin{pmatrix}
		\alpha_{11} & 0 & 0 &0 \\
		0 & \alpha_{22} & \alpha_{23} & 0 \\
		0 & \alpha_{32} & \alpha_{33} & 0 \\
		0 & 0 & 0 & 0
	\end{pmatrix},
	\label{MR}
\end{equation}
with
\begin{align}
	\alpha_{11} = \tau\varrho_{22}+\tau\varrho_{33}&, \quad
	\alpha_{22} = \alpha_{3,3}=-(\tau-1)\varrho_{22},\\
	\alpha_{23} = (1-\tau)\varrho_{23}&, \quad
	\alpha_{32} = (1-\tau)\varrho_{32}.\nonumber
\end{align}
Using Equations ~(\ref{eq:SAtoB}) and ~(\ref{eq:SBtoA}), the quantum steering can be expressed as 
\begin{equation}
	\mathcal{S}_{A \to B}^{AD} = \max\left(0,\frac{\mathcal{N}_{AB}^{AD}-2}{4}\right), \quad \mathcal{S}_{B \to A}^{AD} = \max\left(0,\frac{\mathcal{N}_{BA}^{AD}-2}{4}\right),
	\label{eq:SAtoBAD}
\end{equation}
where \begin{equation}
	\mathcal{N}_{AB}^{AD} = \frac{1}{2}\sum_{i=1}^{4}
	\left[
	\mathcal{I}^{AD}_{x_i, AB}\log_2 \mathcal{I}^{AD}_{x_i, AB}
	+ \mathcal{I}^{AD}_{y_i, AB}\log_2 \mathcal{I}^{AD}_{y_i, AB}
	+ \mathcal{I}^{AD}_{z_i, AB}\log_2 \mathcal{I}^{AD}_{z_i, AB}
	\right]
	- \sum_{j=x,y,z}\sum_{k}
	\mathcal{I}^{AD}_{j_k, A}\log_2 \mathcal{I}^{AD}_{j_k, A},
	\label{eq:IAB}
\end{equation}
\begin{equation}
	\mathcal{N}_{BA}^{AD} = \frac{1}{2}\sum_{i=1}^{4}
	\left[
	\mathcal{I}^{AD}_{x_i, AB}\log_2 \mathcal{I}^{AD}_{x_i, AB}
	+ \mathcal{I}^{AD}_{y_i, AB}\log_2 \mathcal{I}^{AD}_{y_i, AB}
	+ \mathcal{I}^{AD}_{z_i, AB}\log_2 \mathcal{I}^{AD}_{z_i, AB}
	\right]
	- \sum_{j=x,y,z}\sum_{k}
	\mathcal{I}^{AD}_{j_k, B}\log_2 \mathcal{I}^{AD}_{j_k, B},
	\label{eq:IAB}
\end{equation}
and
\begin{align}
	\mathcal{I}^{AD}_{x_{1,2}, AB} &= 1 + 2\,\mathrm{Re}(\alpha_{23}), \qquad  \mathcal{I}^{AD}_{x_{1,2},A} = \mathcal{I}^{AD}_{y_{1,2},A}=1, \nonumber\\
	\mathcal{I}^{AD}_{x_{3,4}, AB} &= 1 - 2\,\mathrm{Re}(\alpha_{23}), \qquad \mathcal{I}^{AD}_{x_{1,2},B} = \mathcal{I}^{AD}_{y_{1,2},B}=1,\\
	\mathcal{I}^{AD}_{y_{1,2}, AB} &= 1 + 2\,\mathrm{Re}(\alpha_{23}),  \qquad\mathcal{I}^{AD}_{z_{1,2},A} = 1 \pm (\alpha_{11}+\alpha_{22}-\alpha_{33}),\nonumber\\
	\mathcal{I}^{AD}_{y_{3,4}, AB} &= 1 - 2\,\mathrm{Re}(\alpha_{23}),\qquad\mathcal{I}^{AD}_{z_{1,2},B} = 1 \pm (\alpha_{11}-\alpha_{22}+\alpha_{33}), \nonumber\\
	\mathcal{I}^{AD}_{z_i, AB} &= 4\alpha_{ii}.\nonumber
\end{align}
Furthermore, an explicit analytical expression for the entanglement negativity, as defined in Eq. ~(\ref{eq:LN_qubits}), can be obtained
\begin{equation}
		\mathcal{L}_{N}({\varrho}^{AD})=\max\left\{0,-2h^{AD}_{\min}\right\},
	\end{equation}
where
\begin{equation}
	h^{\text{AD}}_{\min} = \min \{ \mu_1^{AD}, \mu_2^{AD}, \mu_3^{AD}, \mu_4^{AD} \},
\end{equation}

where \( \mu_{i}^{AD} \) (\(i=1,2,3,4\)) denote the eigenvalues of the partial transpose of the matrix (\ref{MR}). These eigenvalues can be explicitly written as

\begin{align}
	\mu_{1,2}^{AD} &= \frac{\alpha_{11} }{2} \ \pm \ \frac{1}{2} \sqrt{ (\alpha_{11})^2 + 4|\alpha_{23}|^2 }, \nonumber\\
\mu_{3,4}^{AD} &= \frac{\alpha_{22} + \alpha_{33}}{2} \ \pm \ \frac{1}{2} \sqrt{ (\alpha_{22}-\alpha_{33})^2  }.
\end{align}
The quantum coherence for the  state ${\varrho}^{AD}$ is defined as
\begin{equation}
	\mathcal{C}_{l_1}({\varrho^{AD}}) = \lvert \alpha_{23}\rvert+\lvert \alpha_{32}\rvert.
\end{equation}

\subsection{Phase Flip Channel}

The phase flip (PF) channel models stochastic phase errors resulting from random applications of the Pauli operator $\sigma_z$, while leaving the population terms unchanged. The corresponding Kraus operators are
\begin{equation}
	{K}_1=
	\begin{pmatrix}
		\sqrt{\tau} & 0 \\
		0 & \sqrt{\tau}
	\end{pmatrix},
	\qquad
	{K}_2=
	\begin{pmatrix}
		\sqrt{1-\tau} & 0 \\
		0 & -\sqrt{1-\tau}
	\end{pmatrix}.
\end{equation}

The evolved density matrix is given by
\begin{equation}
	{\varrho}^{PF}=
	\begin{pmatrix}
		0 & 0 & 0 & 0 \\
		0 & \varrho_{22} & \beta_{23} & 0 \\
		0 & \beta_{32} & \varrho_{33} & 0 \\
	0 & 0 & 0 & 0
	\end{pmatrix},
\end{equation}
where
\begin{equation}
	\beta_{23}=\varrho_{23}(1-2\tau)^2,
	\qquad
	\beta_{32}=\varrho_{32}(1-2\tau)^2.
\end{equation}
Employing ~(\ref{eq:SAtoB}) and ~(\ref{eq:SBtoA}), we derive the expression for the  the quantum steering
\begin{equation}
	\mathcal{S}_{A \to B}^{PF} = \max\left(0,\frac{\mathcal{N}_{AB}^{PF}-2}{4}\right), \quad \mathcal{S}_{B \to A}^{PF} = \max\left(0,\frac{\mathcal{N}_{BA}^{PF}-2}{4}\right),
	\label{eq:SAtoBPF}
\end{equation}
where \begin{equation}
	\mathcal{N}_{AB}^{PF} = \frac{1}{2}\sum_{i=1}^{4}
	\left[
	\mathcal{I}^{PF}_{x_i, AB}\log_2 \mathcal{I}^{PF}_{x_i, AB}
	+ \mathcal{I}^{PF}_{y_i, AB}\log_2 \mathcal{I}^{PF}_{y_i, AB}
	+ \mathcal{I}^{PF}_{z_i, AB}\log_2 \mathcal{I}^{PF}_{z_i, AB}
	\right]
	- \sum_{j=x,y,z}\sum_{k}
	\mathcal{I}^{PF}_{j_k, A}\log_2 \mathcal{I}^{PF}_{j_k, A},
	\label{eq:IAB}
\end{equation}
\begin{equation}
	\mathcal{N}_{BA}^{PF} = \frac{1}{2}\sum_{i=1}^{4}
	\left[
	\mathcal{I}^{PF}_{x_i, AB}\log_2 \mathcal{I}^{PF}_{x_i, AB}
	+ \mathcal{I}^{PF}_{y_i, AB}\log_2 \mathcal{I}^{PF}_{y_i, AB}
	+ \mathcal{I}^{PF}_{z_i, AB}\log_2 \mathcal{I}^{PF}_{z_i, AB}
	\right]
	- \sum_{j=x,y,z}\sum_{k}
	\mathcal{I}^{PF}_{j_k, B}\log_2 \mathcal{I}^{PF}_{j_k, B},
	\label{eq:IAB}
\end{equation}
and
\begin{align}
	\mathcal{I}^{PF}_{x_{1,2}, AB} &= 1 + 2\,\mathrm{Re}(\beta_{23}), \qquad  \mathcal{I}^{PF}_{x_{1,2},A} = \mathcal{I}^{PF}_{y_{1,2},A}=1, \nonumber\\
	\mathcal{I}^{PF}_{x_{3,4}, AB} &= 1 - 2\,\mathrm{Re}(\beta_{23}), \qquad \mathcal{I}^{PF}_{x_{1,2},B} = \mathcal{I}^{PF}_{y_{1,2},B}=1,\\
	\mathcal{I}^{PF}_{y_{1,2}, AB} &= 1 + 2\,\mathrm{Re}(\beta_{23}),  \qquad\mathcal{I}^{PF}_{z_{1,2},A} = 1, \pm (\varrho_{22}-\varrho_{33}),\nonumber\\
	\mathcal{I}^{PF}_{y_{3,4}, AB} &= 1 - 2\,\mathrm{Re}(\beta_{23}),\qquad\mathcal{I}^{PF}_{z_{1,2},B} = 1 \pm (-\varrho_{22}+\varrho_{33}), \nonumber\\
	\mathcal{I}^{PF}_{z_i, AB} &= 4\varrho_{ii}.\nonumber
\end{align}
Moreover, one can derive an explicit analytical expression for the entanglement negativity, as defined in Eq.~(\ref{eq:LN_qubits})
\begin{equation}
	\mathcal{L}_{N}({\varrho}^{PF})=\max\left\{0,-2h^{PF}_{\min}\right\},
\end{equation}
where
\begin{equation}
	h^{\text{PF}}_{\min} = \min \{ \mu_1^{PF}, \mu_2^{PF}, \mu_3^{PF}, \mu_4^{PF} \},
\end{equation}

\begin{align}
	\mu_{1,2}^{PF} &=  \pm \ \frac{1}{2} \sqrt{  4|\beta_{23}|^2 }, \nonumber\\
	\mu_{3,4}^{PF} &= \frac{\varrho_{22} + \varrho_{33}}{2} \ \pm \ \frac{1}{2} \sqrt{ (\varrho_{22}-\varrho_{33})^2  }.
\end{align}
The quantum coherence of the state ${\varrho}^{PF}$ can be quantified using the $l_1$-norm, which is given by
\begin{equation}
	\mathcal{C}_{l_1}({\varrho^{PF}}) = |\beta_{23}| + |\beta_{32}|.
\end{equation}

\subsection{Phase Damping Channel}

The phase damping (PD) channel describes pure dephasing processes that suppress quantum coherence without affecting the energy populations of the system. The Kraus operators are
\begin{equation}
	{K}_1=
	\begin{pmatrix}
		1 & 0 \\
		0 & \sqrt{1-\tau}
	\end{pmatrix},
	\qquad
	\hat{K}_2=
	\begin{pmatrix}
		0 & 0 \\
		0 & \sqrt{\tau}
	\end{pmatrix}.
\end{equation}

The density matrix under PD noise reads
\begin{equation}
	{\varrho}^{PD}=
	\begin{pmatrix}
		0 & 0 & 0 & 0 \\
		0 & \varrho_{22} & \gamma_{23} & 0 \\
		0 & \gamma_{32} & \varrho_{33} & 0 \\
		0 & 0 & 0 & 0
	\end{pmatrix},\label{122}
\end{equation}
with
\begin{equation}
	\gamma_{23}=\varrho_{23}(1-\tau),
	\qquad
	\gamma_{32}=\varrho_{32}(1-\tau).
\end{equation}
From Eqs.~(\ref{eq:SAtoB}) and~(\ref{eq:SBtoA}), one obtains the following expression for the quantum steering
\begin{equation}
	\mathcal{S}_{A \to B}^{PD} = \max\left(0,\frac{\mathcal{N}_{AB}^{PD}-2}{4}\right), \quad \mathcal{S}_{B \to A}^{PD} = \max\left(0,\frac{\mathcal{N}_{BA}^{PD}-2}{4}\right),
	\label{eq:SAtoBPF}
\end{equation}
where \begin{equation}
	\mathcal{N}_{AB}^{PD} = \frac{1}{2}\sum_{i=1}^{4}
	\left[
	\mathcal{I}^{PD}_{x_i, AB}\log_2 \mathcal{I}^{PD}_{x_i, AB}
	+ \mathcal{I}^{PD}_{y_i, AB}\log_2 \mathcal{I}^{PD}_{y_i, AB}
	+ \mathcal{I}^{PD}_{z_i, AB}\log_2 \mathcal{I}^{PD}_{z_i, AB}
	\right]
	- \sum_{j=x,y,z}\sum_{k}
	\mathcal{I}^{PD}_{j_k, A}\log_2 \mathcal{I}^{PD}_{j_k, A},
	\label{eq:IAB}
\end{equation}
\begin{equation}
	\mathcal{N}_{BA}^{PD} = \frac{1}{2}\sum_{i=1}^{4}
	\left[
	\mathcal{I}^{PD}_{x_i, AB}\log_2 \mathcal{I}^{PD}_{x_i, AB}
	+ \mathcal{I}^{PD}_{y_i, AB}\log_2 \mathcal{I}^{PD}_{y_i, AB}
	+ \mathcal{I}^{PD}_{z_i, AB}\log_2 \mathcal{I}^{PD}_{z_i, AB}
	\right]
	- \sum_{j=x,y,z}\sum_{k}
	\mathcal{I}^{PD}_{j_k, B}\log_2 \mathcal{I}^{PD}_{j_k, B},
	\label{eq:IAB}
\end{equation}
and
\begin{align}
	\mathcal{I}^{PD}_{x_{1,2}, AB} &= 1 + 2\,\mathrm{Re}(\gamma_{23}), \qquad  \mathcal{I}^{PD}_{x_{1,2},A} = \mathcal{I}^{PD}_{y_{1,2},A}=1, \nonumber\\
	\mathcal{I}^{PD}_{x_{3,4}, AB} &= 1 - 2\,\mathrm{Re}(\gamma_{23}), \qquad \mathcal{I}^{PD}_{x_{1,2},B} = \mathcal{I}^{PD}_{y_{1,2},B}=1,\\
	\mathcal{I}^{PD}_{y_{1,2}, AB} &= 1 + 2\,\mathrm{Re}(\gamma_{23}),  \qquad\mathcal{I}^{PD}_{z_{1,2},A} = 1 \pm (\varrho_{22}-\varrho_{33}),\nonumber\\
	\mathcal{I}^{PD}_{y_{3,4}, AB} &= 1 - 2\,\mathrm{Re}(\gamma_{23}),\qquad\mathcal{I}^{PD}_{z_{1,2},B} = 1 \pm (-\varrho_{22}+\varrho_{33}), \nonumber\\
	\mathcal{I}^{PD}_{z_i, AB} &= 4\varrho_{ii}.\nonumber
\end{align}
Furthermore, a closed-form analytical expression for the entanglement negativity, as defined in Eq.~(\ref{eq:LN_qubits}), can be obtained as
\begin{equation}
	\mathcal{L}_{N}(\varrho^{PD})=\max\left\{0,-2h^{PD}_{\min}\right\},
\end{equation}
where
\begin{equation}
	h^{PD}_{\min}=\min\left\{\mu_1^{PD},\mu_2^{PD},\mu_3^{PD},\mu_4^{PD}\right\}.
\end{equation}
Here, $\mu_i^{PD}$ ($i=1,2,3,4$) denote the eigenvalues of the partial transpose of the density matrix given in Eq.~(\ref{122}). These eigenvalues are explicitly given by
\begin{align}
	\mu_{1,2}^{PD} &= 
	\pm \frac{1}{2}\sqrt{4|\gamma_{23}|^{2}}, \nonumber\\
	\mu_{3,4}^{PD} &= \frac{\varrho_{22}+\varrho_{33}}{2}
	\pm \frac{1}{2}\sqrt{(\varrho_{22}-\varrho_{33})^{2}}.
\end{align}
The quantum coherence of the state $\varrho^{PD}$, quantified by the $l_{1}$-norm, is given by
\begin{equation}
	\mathcal{C}_{l_1}\!\left(\varrho^{PD}\right)
	= \left|\gamma_{23}\right| + \left|\gamma_{32}\right| .
\end{equation}

\section{RESULTS AND DISCUSSIONS}   \label{sec:5}

The numerical results presented in this work are grounded in the experimental data reported by the Daya Bay, KamLAND, MINOS, T2K, and JUNO collaborations \cite{DayaBay2017, KamLAND2005, MINOS2008, T2K2011, JUNO2016}, which provide the foundational oscillation parameters for our model. These experiments explore complementary oscillation regimes, encompassing reactor and accelerator-based neutrino sources across short, medium, and long-baseline configurations \cite{NOvA2026}. The reactor antineutrino experiments Daya Bay and KamLAND focus on electron-antineutrino disappearance at short and medium baselines, respectively; these facilitate high-precision measurements of the mixing angles $\theta_{13}$ and $\theta_{12}$, as well as the corresponding mass-squared differences $\Delta m^2_{ee}$ and $\Delta m^2_{21}$, as summarized in Table~\ref{tab:1}. JUNO, a next-generation medium-baseline reactor experiment, is designed to achieve high sensitivity to the neutrino mass ordering and to further refine solar oscillation parameters through precise measurements of antineutrino energy spectra. In contrast, the accelerator-based long-baseline experiments MINOS and T2K investigate muon-neutrino disappearance and appearance channels at higher energies, providing stringent constraints on the atmospheric mass-squared difference $\Delta m^2_{32}$ and the mixing angle $\theta_{23}$. Notably, T2K has reported strong evidence for electron-neutrino appearance in a muon-neutrino beam, contributing significantly to the global determination of oscillation parameters. Collectively, the diverse baselines, energy ranges, and detection strategies of these five experiments provide a comprehensive framework for investigating neutrino oscillation phenomena and their associated quantum informational properties. The experimental parameters used throughout this work are summarized in Table~\ref{tab:1}.

\subsection{Without dephasing effect}

\begin{figure}[H]
	\includegraphics[scale=0.58]{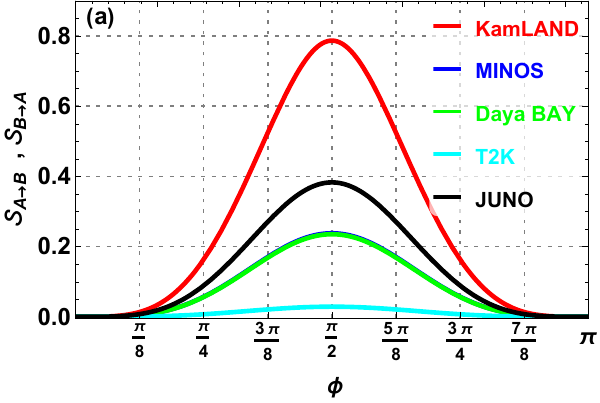}
	\includegraphics[scale=0.58]{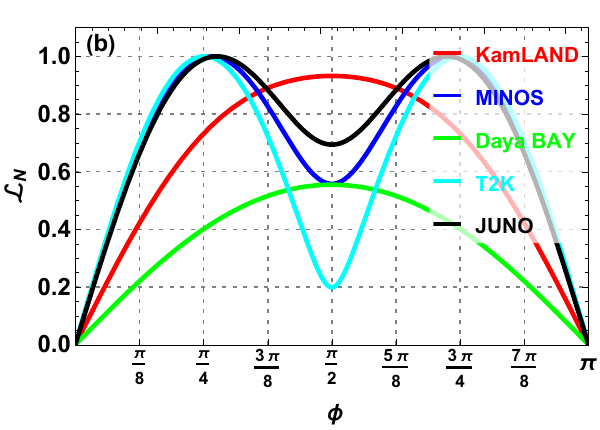}
		\includegraphics[scale=0.58]{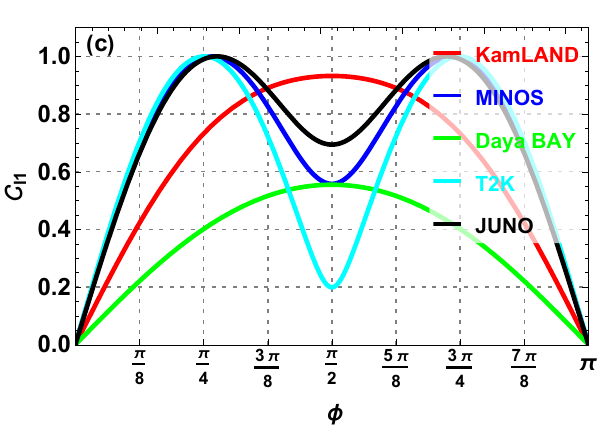}
	\caption{Plot of the quantum steering, logarithmic negativity, and quantum coherence as a function of $\phi$ for the various experiments presented in Table~\ref{tab:1} }
	\label{fig:1}
\end{figure}

To assess how the parameter $\phi$ influences quantum resources, we apply the experimental values from Table~\ref{tab:1} to compute the steering, negativity, and coherence for each experiment. The curves in Fig.~\ref{fig:1}(a-c) correspond to these calculations under the absence of decoherence.
Fig.~\ref{fig:1}(a) shows that quantum steering increases monotonically with the parameter $\phi$, reaching its maximum at $\phi = \pi/2$, which corresponds to optimal quantum correlations. Beyond this point, the steering decreases symmetrically and vanishes at $\phi = \pi$, reflecting the oscillatory dependence of quantum correlations on $\phi$. The different peak amplitudes observed across the experiments highlight the influence of their specific parameters on the strength of quantum steering. As illustrated in Fig.~\ref{fig:1}(b), the logarithmic negativity $\mathcal{L}_N$ varies with the phase $\phi$ for the KamLAND, MINOS, Daya Bay, T2K, and JUNO experiments. For all cases, $\mathcal{L}_N$ initially increases with $\phi$, reaching a peak at intermediate values. Beyond this point, $\mathcal{L}_N$ gradually decreases, eventually vanishing at $\phi = \pi$. The curves reveal distinct amplitudes and oscillatory patterns characteristic of each experiment; for instance, KamLAND and JUNO show nearly symmetric peaks around $\phi = \pi/2$, whereas T2K exhibits a more pronounced dip near $\phi = \pi/2$. This behavior reflects the phase-dependent evolution of quantum correlations in these neutrino oscillation systems. Finally, as shown in Fig.~\ref{fig:1}(c), quantum coherence exhibits a dynamical trend similar to that of the logarithmic negativity.

\subsection{Under dephasing effect}

In this section, we study the evolution of quantum steering, logarithmic negativity, and quantum coherence for the experiments listed in Table~\ref{tab:1}. Their dependence on the decoherence parameter $\tau$ is analyzed across various channels, including amplitude damping (AD), phase damping (PD), and phase flip (PF). 
\begin{figure}[H]
	\includegraphics[scale=0.58]{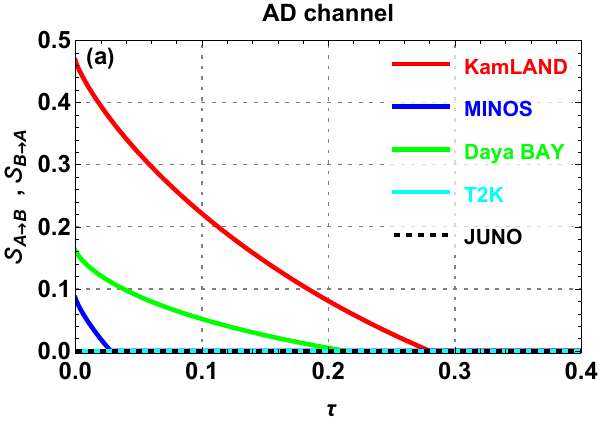}
	\includegraphics[scale=0.58]{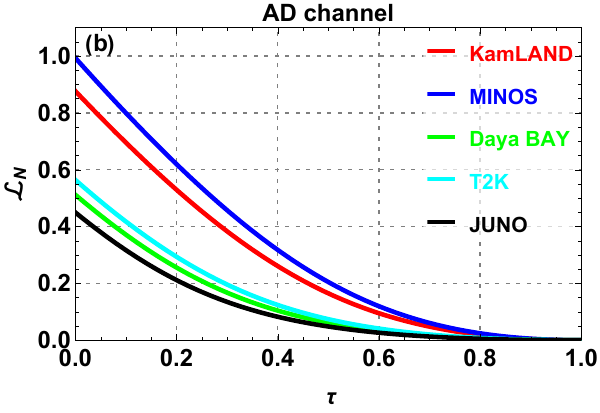}
	\includegraphics[scale=0.58]{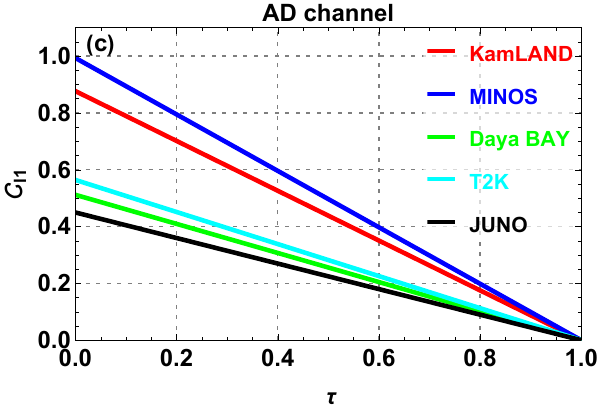}
	\caption{Quantum steering $\mathcal{S}_{A\to B}$ (a), logarithmic negativity $\mathcal{L}_N$ (b), and quantum coherence $\mathcal{C}_{l_1}$ (c) as functions of the decoherence parameter $\tau$ for the AD channel, considering the various experiments listed in Table~\ref{tab:1}.}
\label{fig:2}
\end{figure}

The dependence of quantum steering ($\mathcal{S}_{A\to B}$) on the decoherence parameter $\tau$ under amplitude damping (AD) is shown in Fig.~\ref{fig:2}(a-c) for the experiments listed in Table~\ref{tab:1}. We remark that the quantum steering monotonically decreases with increasing $\tau$, reflecting the progressive loss of quantum correlations induced by environmental noise. As $\tau$ increases further, the steering is completely suppressed and eventually vanishes. 
The decay rate and the robustness of steering depend on the specific experimental parameters, with KamLAND exhibiting the highest resilience, followed by Daya Bay, MINOS, T2K, and JUNO, as shown in Fig.~\ref{fig:2}(c).

Figure~\ref{fig:2}(b) and Figure~\ref{fig:2}(c) illustrate the evolution of logarithmic negativity and quantum coherence under the amplitude damping (AD) channel as a function of the decoherence parameter $\tau$ for various neutrino experiments.
It can be seen that for small decoherence ($\tau \lesssim 0.2$), both entanglement and coherence remain relatively high, indicating the persistence of quantum correlations. As $\tau$ increases ($0.2 \lesssim \tau \lesssim 0.6$), a steady decay is observed in both quantities, reflecting the gradual loss of quantum correlations  due to environmental effects.
For larger decoherence ($\tau \gtrsim 0.6$), logarithmic negativity is almost completely suppressed, signaling the loss of entanglement, while quantum coherence, although significantly reduced, can still retain small residual values, marking a partial persistence of phase information. The rate of decay depends on the experimental setup, with MINOS and KamLAND showing greater resilience across the $\tau$ range, whereas Daya Bay, T2K, and JUNO experience faster degradation.

A clear hierarchy emerges among the different quantum correlations under the amplitude damping channel. Quantum steering vanishes first with increasing decoherence, followed by entanglement, while quantum coherence remains the most robust and persists even after the suppression of both steering and entanglement.

\begin{figure}[H]
	\includegraphics[scale=0.58]{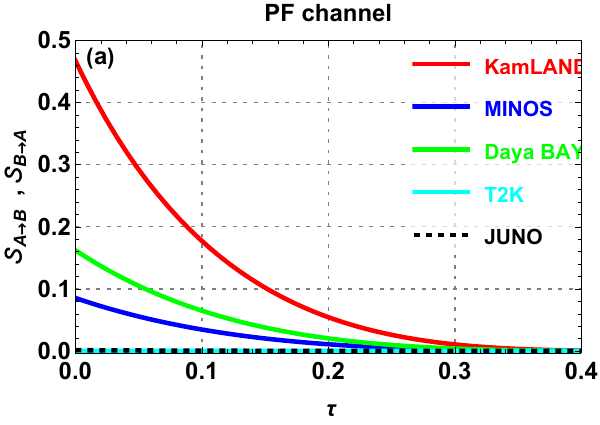}
	\includegraphics[scale=0.58]{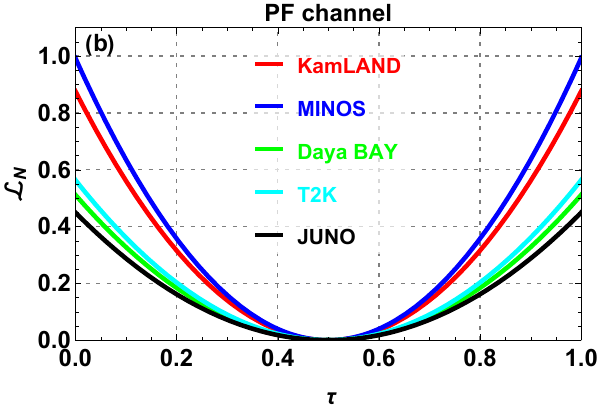}
	\includegraphics[scale=0.58]{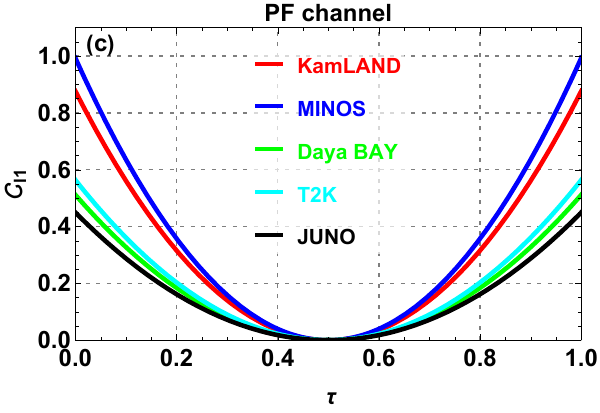}
	\caption{Plot of the quantum steering $\mathcal{S}_{A\to B}$ (a), logarithmic negativity $\mathcal{L}_N$ (b), and quantum coherence $\mathcal{C}_{l1}$ (c) against $\tau$ under PF channel for various experiments presented in Table~\ref{tab:1}.}
	\label{fig:3}
\end{figure}
The influence of the Phase Flip (PF) channel on quantum correlations and coherence is illustrated in Fig.~\ref{fig:3}.
As shown in Fig.~\ref{fig:3}(a), quantum steering $\mathcal{S}_{A\to B}$ decreases monotonically with increasing decoherence parameter $\tau$ for all neutrino experiments and vanishes at large $\tau$, indicating the loss of steerability. One can see that the KamLAND shows the highest robustness against phase noise, while the JUNO is the most sensitive.

A direct comparison between the AD and PF channels shows that, although quantum steering decreases monotonically with $\tau$ in both cases, the decay is faster under amplitude damping.
	This indicates that the AD channel induces a stronger degradation of quantum correlations than the PF channel, which mainly affects phase coherence.

Figures~\ref{fig:3}(b) and \ref{fig:3}(c) demonstrate that logarithmic negativity and quantum coherence exhibit identical dynamical behavior under the PF channel, revealing a direct correspondence between entanglement and coherence in a phase-flip environment. Both quantities undergo a symmetric decay–revival dynamics as a function of $\tau$, reaching a minimum at intermediate decoherence strengths.
This indicates that phase noise affects entanglement and coherence in an equivalent manner, highlighting their intrinsic connection under purely dephasing processes.

Logarithmic negativity and quantum coherence follow identical dynamical curves under both AD and PF channels.
While AD induces a monotonic  decay, the PF channel gives rise to a symmetric decay–revival behavior due to its purely dephasing nature.

\begin{figure}[H]
	\includegraphics[scale=0.58]{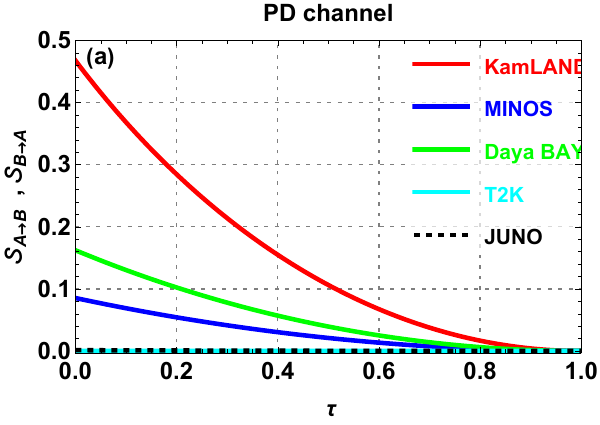}
	\includegraphics[scale=0.58]{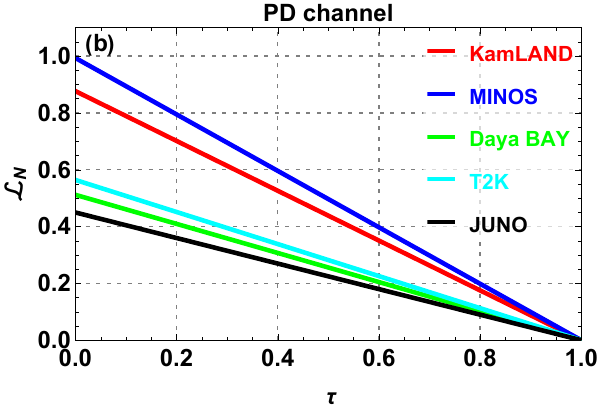}
	\includegraphics[scale=0.58]{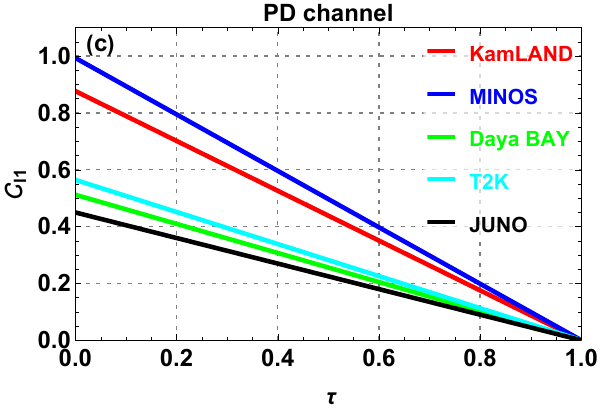}
	\caption{Plot of the quantum steering $\mathcal{S}_{A\to B}$ (a), logarithmic negativity $\mathcal{L}_N$ (b), and quantum coherence $\mathcal{C}_{l1}$ (c) versus $\tau$ under PD channel for various experiments presented in Table~\ref{tab:1}.}
	\label{fig:4}
\end{figure}
The effect of the Phase Damping (PD) channel on quantum resources is illustrated in Fig.~\ref{fig:4}. As shown in Fig.~\ref{fig:4}(a), quantum steering $\mathcal{S}_{A\to B}$ decreases monotonically with the decoherence parameter $\tau$ for all neutrino experiments and vanishes at $\tau \simeq 1$. Also, the KamLAND shows the highest robustness, while the JUNO is the most sensitive to phase damping.

Figures~\ref{fig:4}(b) and \ref{fig:4}(c) show that logarithmic negativity and quantum coherence evolve identically under the PD channel and decrease linearly with increasing $\tau$. MINOS and KamLAND retain larger values over the entire decoherence range, whereas Daya Bay, T2K, and JUNO experience a faster degradation.

A comparison between the different quantum resources indicates that quantum steering is more fragile under the PD channel than entanglement and coherence, as it vanishes at smaller values of $\tau$.
In contrast, logarithmic negativity and quantum coherence exhibit identical and more robust linear decay, highlighting a direct correspondence between these two resources in a phase-damping environment.

\begin{figure}[H]
		\includegraphics[scale=0.58]{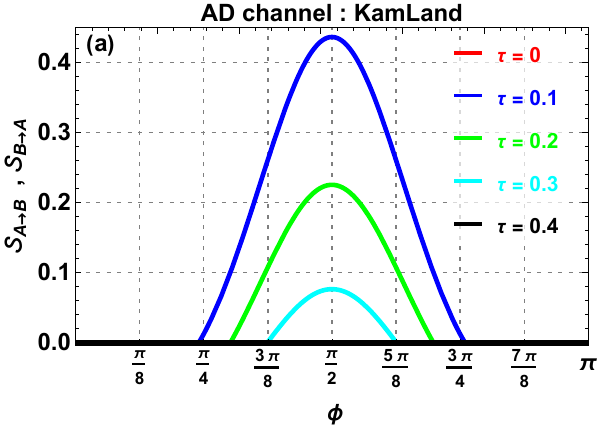}
	\includegraphics[scale=0.58]{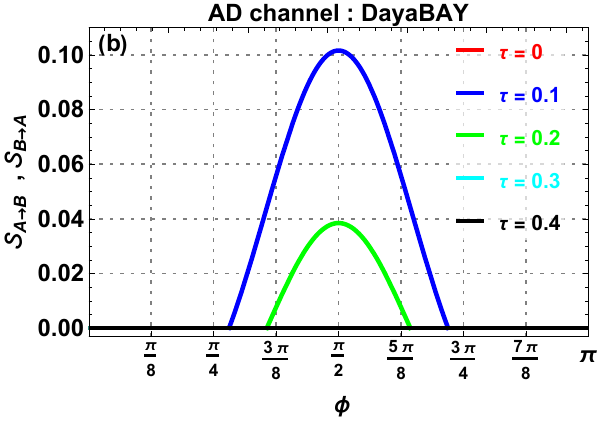}
	\includegraphics[scale=0.58]{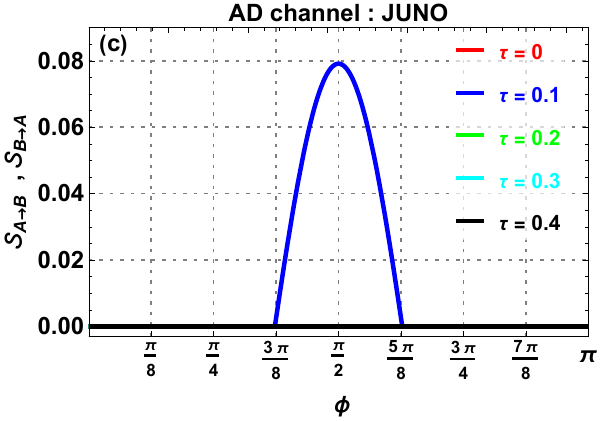}
	\includegraphics[scale=0.58]{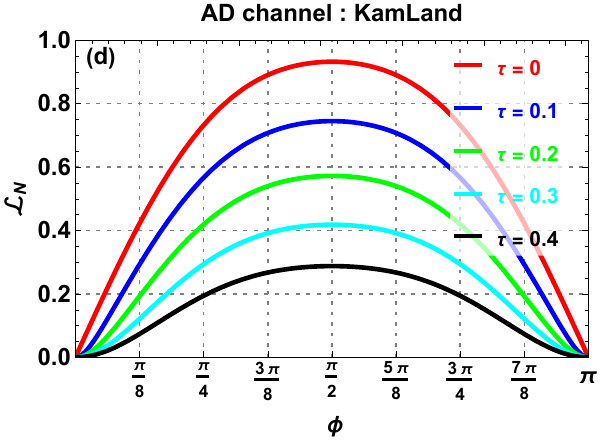}
	\includegraphics[scale=0.58]{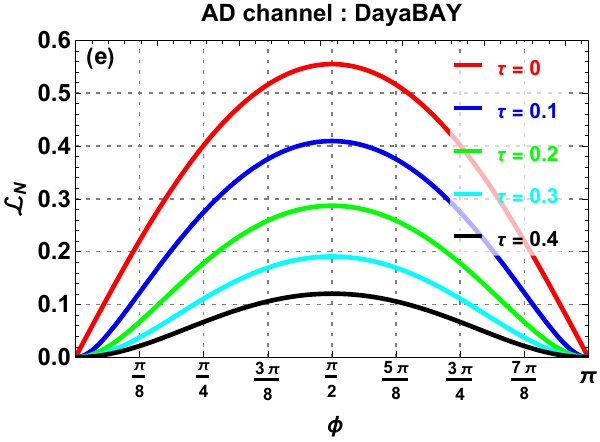}
	\includegraphics[scale=0.58]{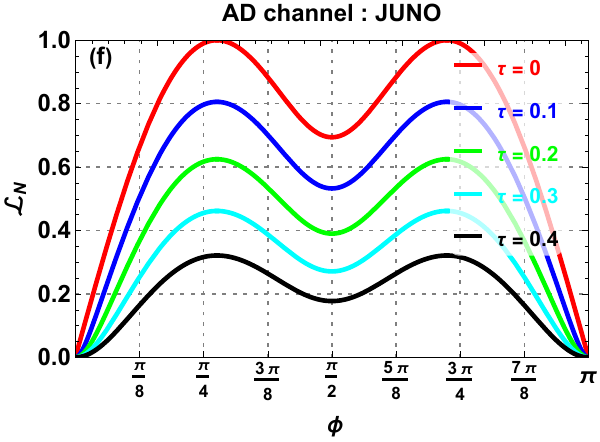}
		\includegraphics[scale=0.58]{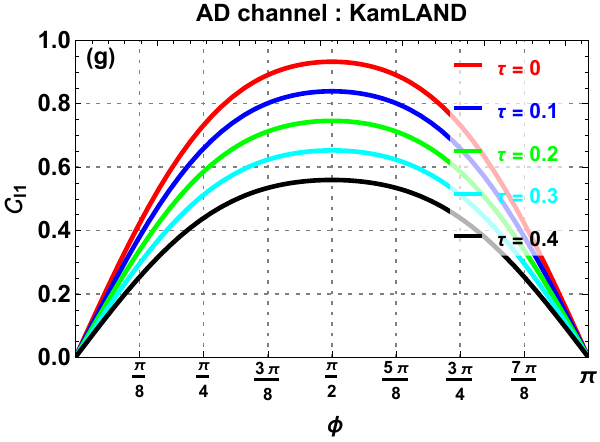}
	\includegraphics[scale=0.58]{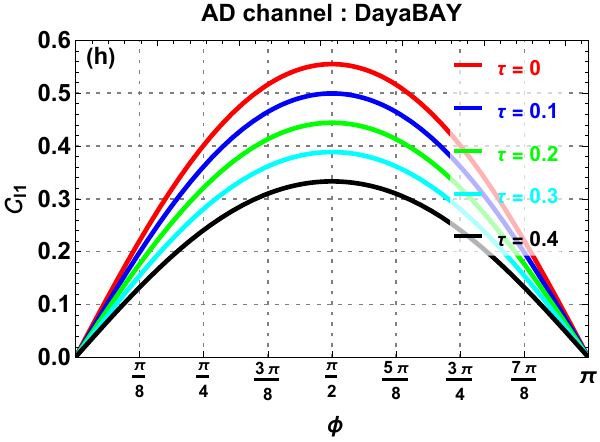}
	\includegraphics[scale=0.58]{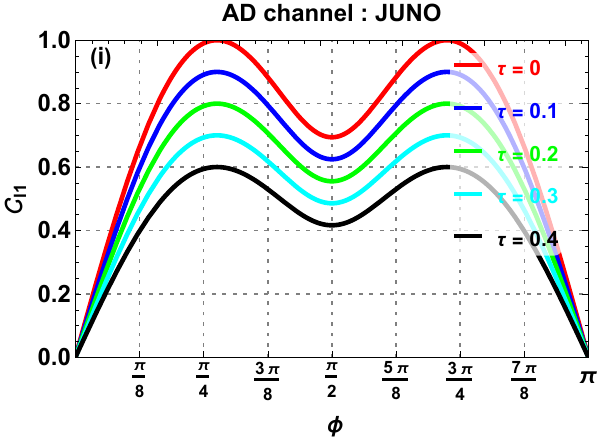}
	\caption{Plot of quantum steering $\mathcal{S}_{A\to B}$ (a-c), logarithmic negativity $\mathcal{L}_N$ (d-f), and quantum coherence $\mathcal{C}_{l1}$ (g-i) as functions of the parameter $\phi$ under the amplitude damping (AD) channel for the KamLAND, Daya Bay, and JUNO experiments, shown for different values of the decoherence parameter $\tau$.}
	\label{fig:5}
\end{figure}

Figure~\ref{fig:5} illustrates the dependence of quantum steering, logarithmic negativity, and quantum coherence on the parameter $\phi$ under the amplitude damping (AD) channel for the KamLAND, Daya Bay, and JUNO experiments. Panels (a)--(c) show the behavior of quantum steering $\mathcal{S}_{A\to B}$ for increasing values of the decoherence parameter $\tau$. We remark that for all experiments, steering vanishes in the vicinity of $\phi = 0$ and $\phi = \pi$, while reaching its maximum around $\phi = \pi/2$. Moreover, as $\tau$ increases, the amplitude of the steering curves is progressively reduced, indicating the strong sensitivity of steerability to dissipative effects induced by the AD channel. Among the considered setups, KamLAND displays the highest steering magnitude, whereas JUNO shows a more restricted steerable region.

The middle panels Fig.~\ref{fig:5}(d-f) depict the evolution of logarithmic negativity as a function of $\phi$.
For KamLAND and Daya Bay, entanglement exhibits a smooth and symmetric profile with a single maximum located around $\phi=\pi/2$.
In contrast, for JUNO the entanglement distribution develops a bimodal structure, characterized by two symmetric maxima around $\phi=\pi/2$, separated by a local minimum at $\phi=\pi/2$

Furthermore, increasing decoherence leads to a systematic suppression of logarithmic negativity, although nonzero entanglement persists for moderate values of $\tau$, particularly for KamLAND and Daya Bay.

The lower panels Fig.~\ref{fig:5}(g-i) present the corresponding behavior of quantum coherence. Similar to entanglement, coherence exhibits a strong dependence on $\phi$ and decreases monotonically as the decoherence parameter increases. While the overall qualitative behavior is comparable across the three experiments, the reduction rate depends on the experimental configuration, with JUNO experiencing the most pronounced degradation. These results highlight the role of the parameter $\phi$ in shaping quantum resources and demonstrate the detrimental impact of amplitude damping on steering, entanglement, and coherence.

Overall, a clear hierarchy among the quantum resources is observed. 
Entanglement, quantified by the logarithmic negativity, is the most robust against decoherence, 
followed by quantum coherence, while quantum steering is the most fragile resource. 
For all configurations, the magnitude of these resources decreases monotonically with increasing noise strength, 
preserving the same qualitative ordering.

\begin{figure}[H]
	\includegraphics[scale=0.58]{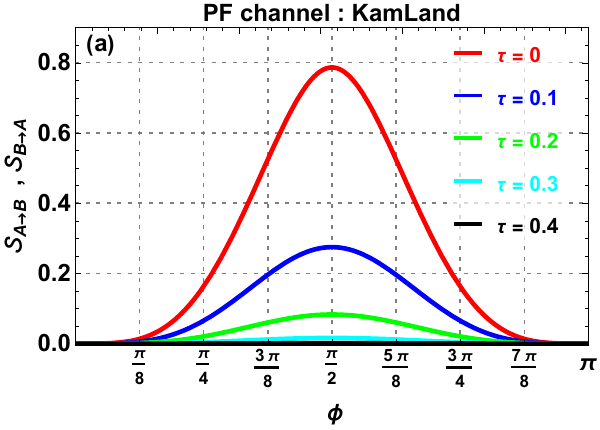}
	\includegraphics[scale=0.58]{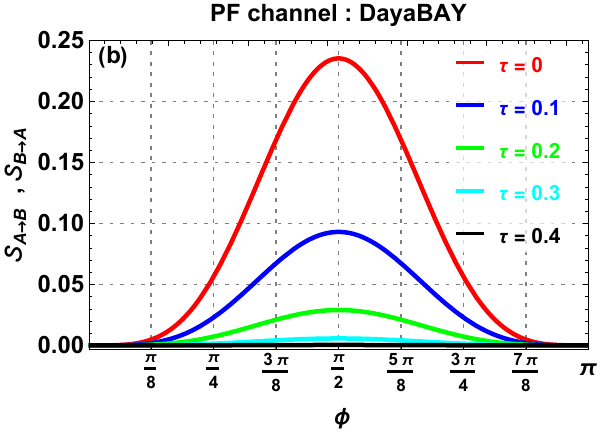}
	\includegraphics[scale=0.58]{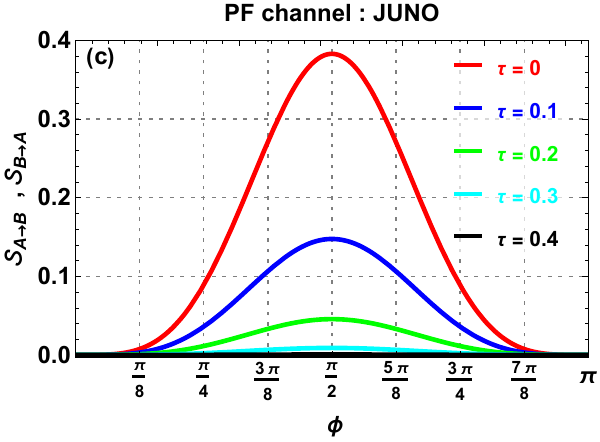}
	\includegraphics[scale=0.58]{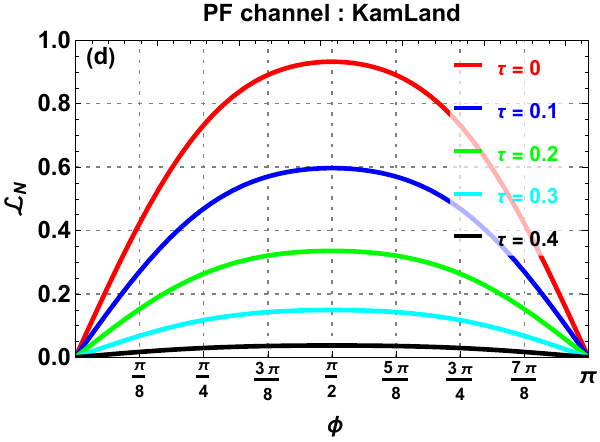}
	\includegraphics[scale=0.58]{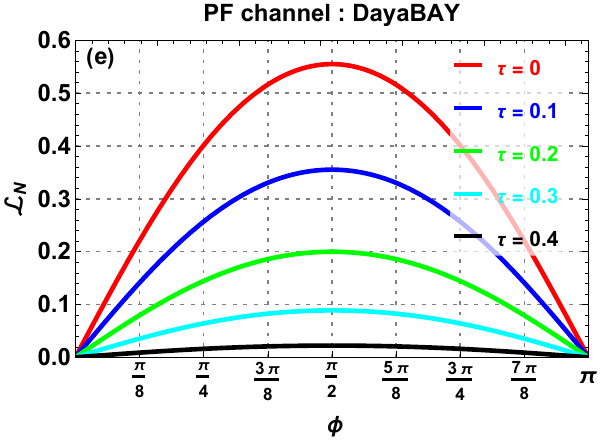}
	\includegraphics[scale=0.58]{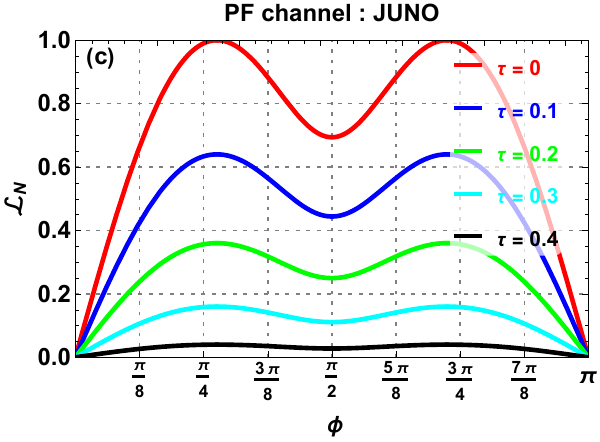}
	\includegraphics[scale=0.58]{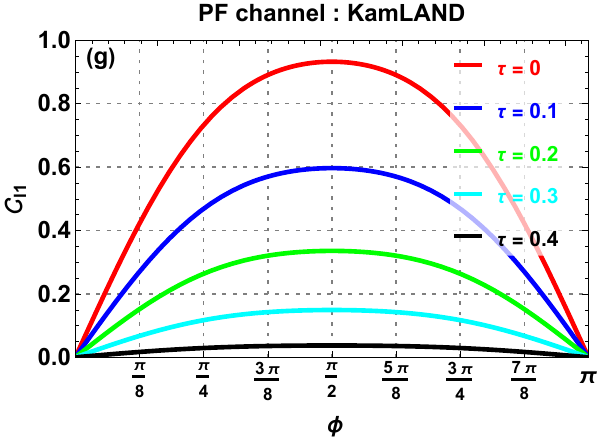}
	\includegraphics[scale=0.58]{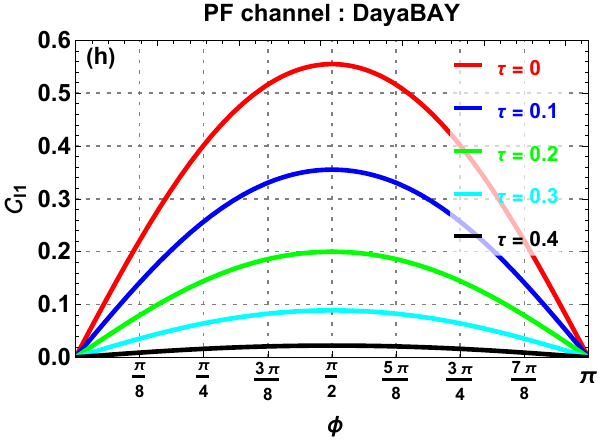}
	\includegraphics[scale=0.58]{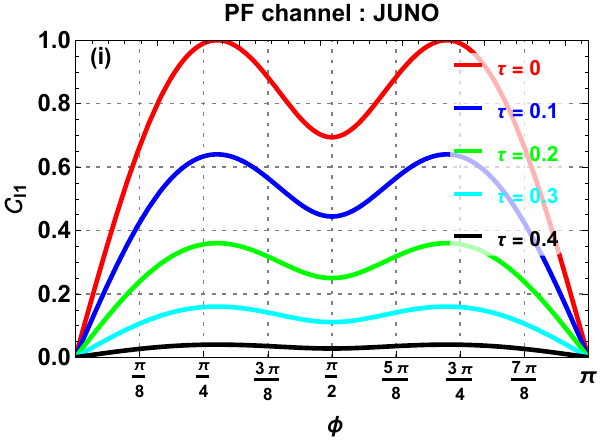}
	\caption{Plot of quantum steering $\mathcal{S}_{A\to B}$ (a-c), logarithmic negativity $\mathcal{L}_N$ (d-f), and quantum coherence $\mathcal{C}_{l1}$ as functions of the parameter $\phi$ under the phase flip (PF) channel for the KamLAND, Daya Bay, and JUNO experiments for different values of the decoherence parameter $\tau$.}
	\label{fig:6}
\end{figure}

Figure~\ref{fig:6} presents the behavior of quantum steering, logarithmic negativity, and quantum coherence as functions of the parameter $\phi$ under the Phase Flip (PF) channel for the KamLAND, Daya Bay, and JUNO experiments. Panels (a)-(c) illustrate the variation of quantum steering $\mathcal{S}_{A\to B}$ for several values of the decoherence parameter $\tau$. In all configurations, steering is suppressed in the vicinity of $\phi = 0$ and $\phi = \pi$, while attaining its largest values around $\phi = \pi/2$. Increasing $\tau$ leads to a gradual narrowing and attenuation of the steering profiles, revealing the sensitivity of steerability to phase-flip noise. KamLAND maintains the strongest steering response, whereas JUNO exhibits a reduced steerable domain.

The middle panels Fig.~\ref{fig:5}(d-f) show the dependence of logarithmic negativity on $\phi$. Entanglement exhibits a symmetric structure with respect to $\phi = \pi/2$ for all experiments. As the decoherence parameter increases, logarithmic negativity is progressively diminished, indicating a weakening of entanglement under the PF channel, although finite values remain observable for moderate $\tau$, particularly in the KamLAND and Daya Bay cases.

The lower panels Fig.~\ref{fig:5}(g-i) display the evolution of quantum coherence with $\phi$. Coherence varies smoothly with $\phi$ and decreases monotonically as decoherence strengthens. The overall behavior is similar across the three experiments, but the magnitude of coherence is more strongly reduced for JUNO. These results demonstrate how phase-flip noise alters the $\phi$-dependent structure of steering, entanglement, and coherence, while preserving their symmetric features.

\begin{figure}[H]
	\includegraphics[scale=0.58]{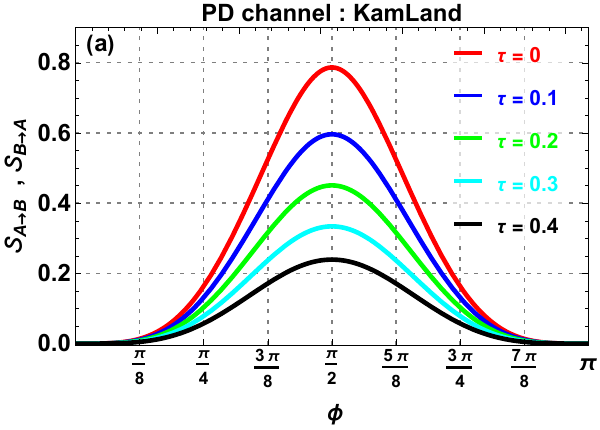}
	\includegraphics[scale=0.58]{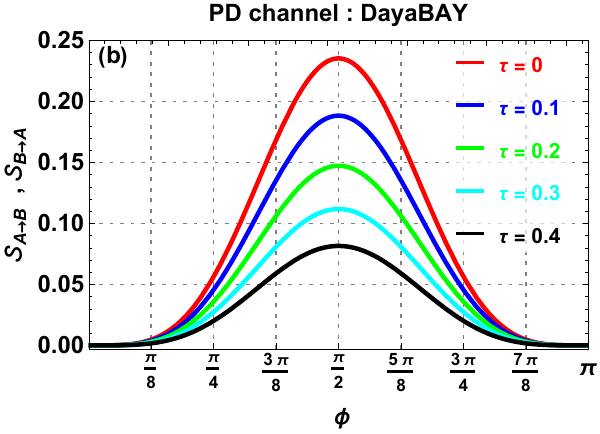}
	\includegraphics[scale=0.58]{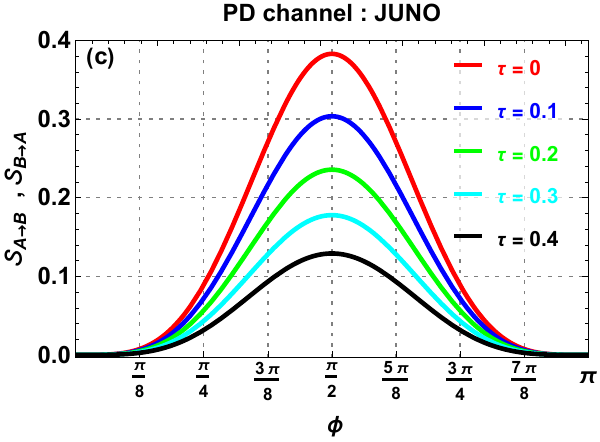}
	\includegraphics[scale=0.58]{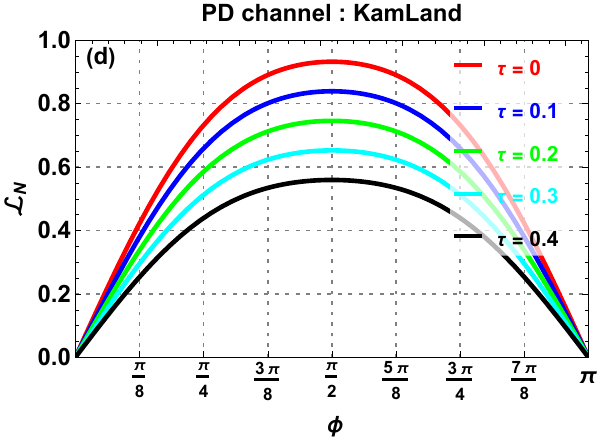}
	\includegraphics[scale=0.58]{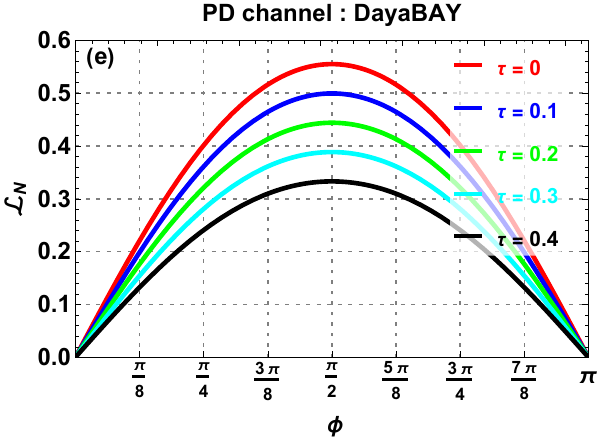}
	\includegraphics[scale=0.58]{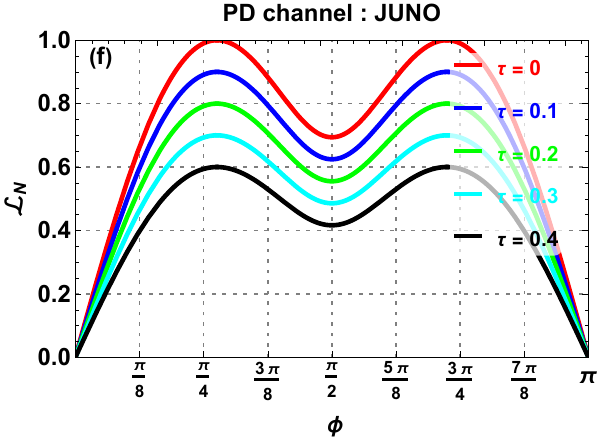}
	\includegraphics[scale=0.58]{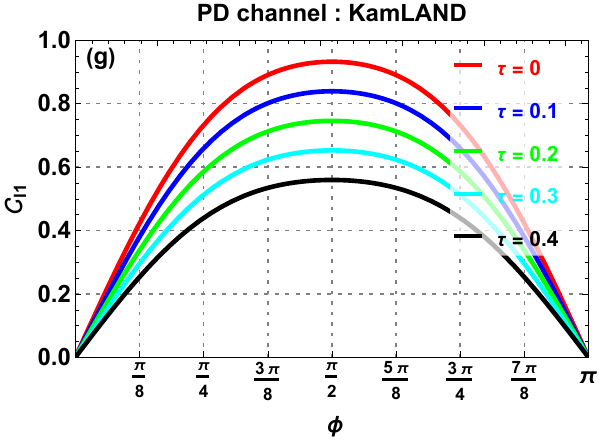}
	\includegraphics[scale=0.58]{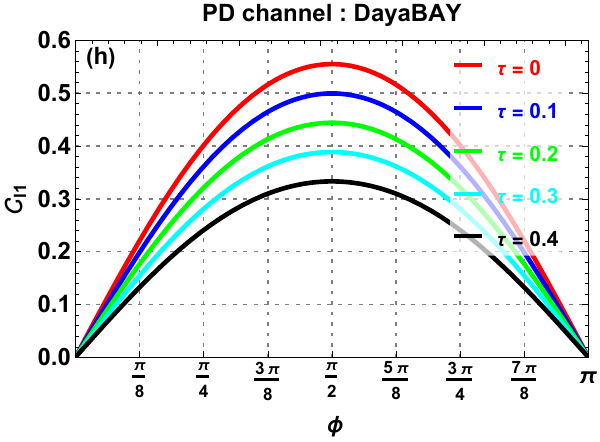}
	\includegraphics[scale=0.58]{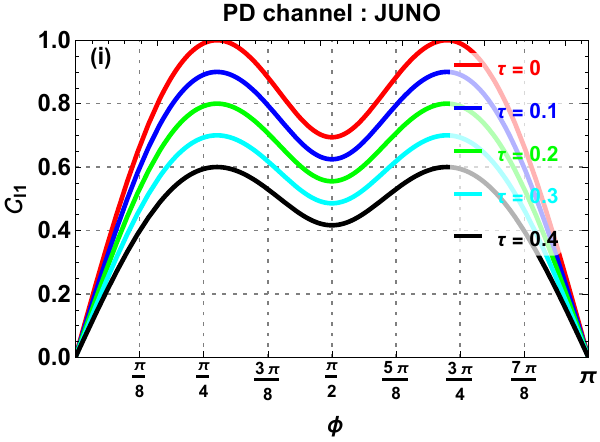}
	\caption{Plot of quantum steering $\mathcal{S}_{A\to B}$, logarithmic negativity $\mathcal{L}_N$, and quantum coherence $\mathcal{C}_{l1}$ as functions of the parameter $\phi$ under the phase damping (PD) channel for the KamLAND, Daya Bay, and JUNO experiments, shown for different values of  $\tau$.}
	\label{fig:7}
\end{figure}
Figure~\ref{fig:7} displays the evolution of quantum steering, logarithmic negativity, and quantum coherence as functions of the parameter $\phi$ under the Phase Damping (PD) channel for the KamLAND, Daya Bay, and JUNO experiments. Quantum steering, shown in panels (a)--(c), is suppressed near $\phi = 0$ and $\phi = \pi$ and attains its maximum around $\phi = \pi/2$, while its magnitude decreases steadily as the decoherence parameter $\tau$ increases. The middle and lower panels show that logarithmic negativity and quantum coherence exhibit smooth and symmetric profiles with respect to $\phi$, both decreasing monotonically with increasing $\tau$. KamLAND and Daya Bay preserve larger values of these quantities compared to JUNO. Overall, the results indicate that phase damping progressively reduces steering, entanglement, and coherence without altering their $\phi$-dependent structure.

A comparative analysis of the amplitude damping (AD), phase-flip (PF), and phase-damping (PD) channels reveals distinct effects on quantum resources. 
The AD channel leads to the strongest suppression of quantum steering, entanglement, and coherence, as evidenced by the rapid reduction of their amplitudes with increasing noise strength. 
In contrast, the PF and PD channels mainly affect phase-dependent quantum correlations, resulting in a more gradual degradation of quantum resources while preserving the symmetry of their $\phi$-dependent profiles. 
Although all channels reduce the magnitude of quantum correlations as the decoherence parameter increases, PF and PD allow nonzero entanglement and coherence to persist over a wider range of noise strengths. 
Moreover, characteristic phase-dependent features, such as the bimodal entanglement structure observed in the JUNO configuration, remain clearly visible under PF and PD, whereas they are more strongly suppressed under the AD channel.

\begin{figure}[H]
	\includegraphics[scale=0.58]{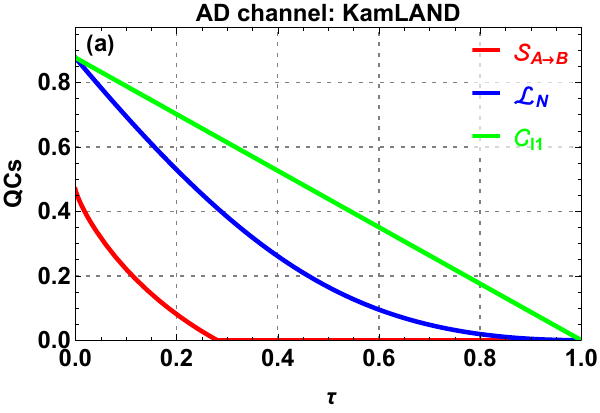}
	\includegraphics[scale=0.58]{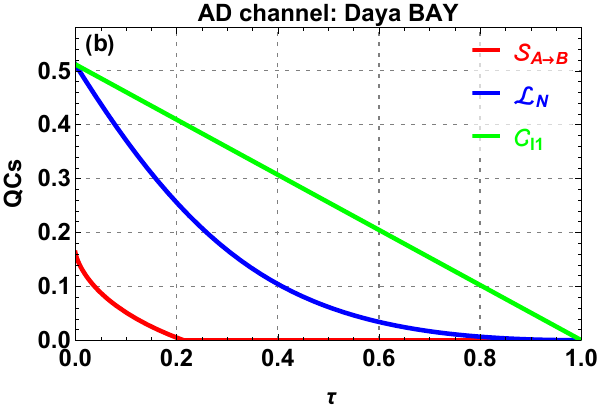}
	\includegraphics[scale=0.58]{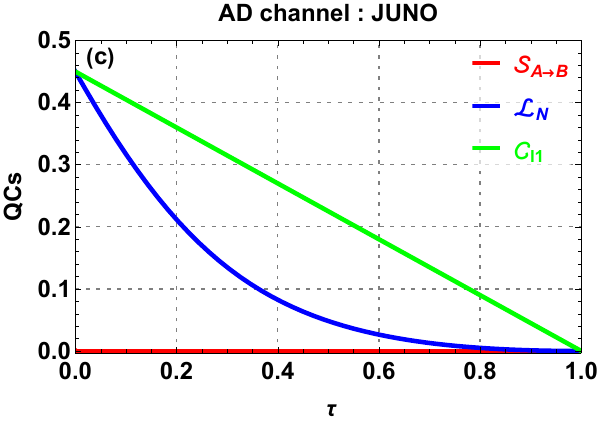}
	\includegraphics[scale=0.58]{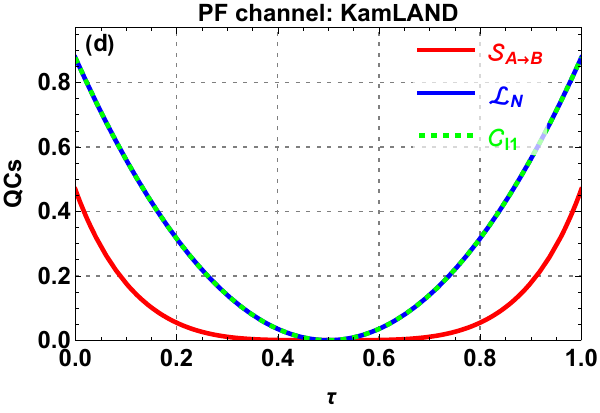}
	\includegraphics[scale=0.58]{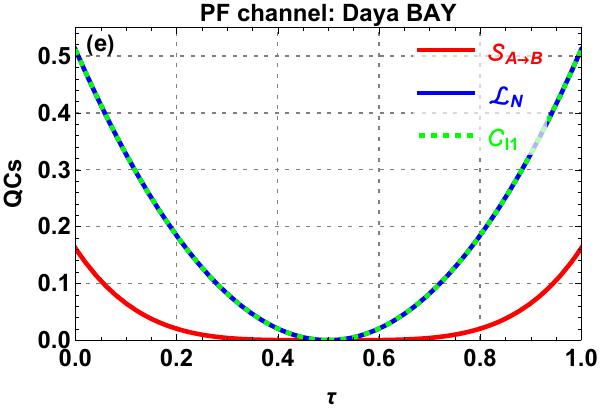}
	\includegraphics[scale=0.58]{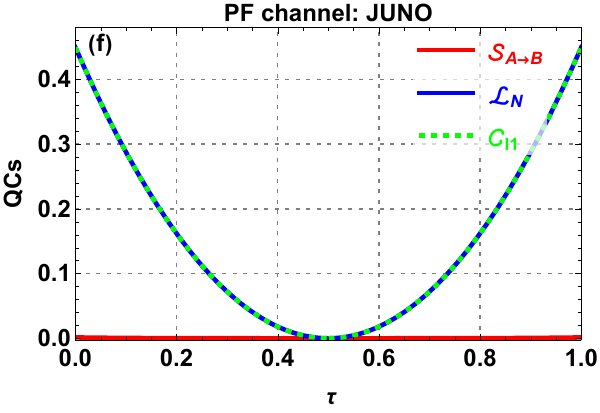}
	\includegraphics[scale=0.58]{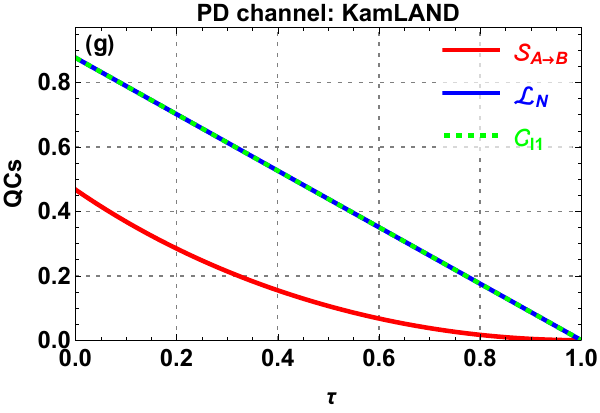}
	\includegraphics[scale=0.58]{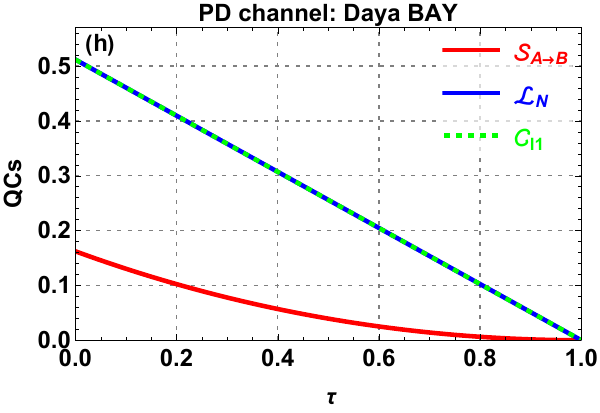}
	\includegraphics[scale=0.58]{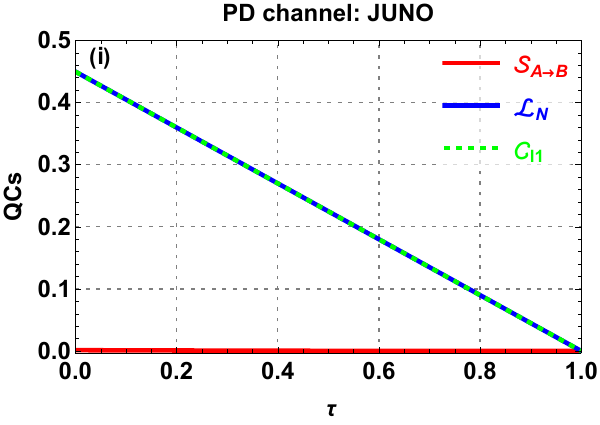}
	\caption{Comparison of quantum steering, logarithmic negativity, and quantum coherence under the amplitude damping (AD), phase flip (PF), and phase damping (PD) channels for the KamLAND, Daya Bay, and JUNO experiments.}
	\label{fig:8}
\end{figure}

Figure~\ref{fig:8} highlights a clear hierarchy in the robustness of quantum correlations under the amplitude damping (AD) channel for the KamLAND, Daya Bay, and JUNO experiments. While quantum steering $\mathcal{S}_{A\to B}$ undergoes a rapid decay and disappears at relatively small values of the decoherence parameter $\tau$, entanglement quantified by the logarithmic negativity exhibits a significantly slower degradation and survives up to larger $\tau$. In contrast, quantum coherence shows the highest resilience against decoherence, persisting over the entire range of $\tau$ even when entanglement has nearly vanished. This behavior indicates that coherence is intrinsically more robust than logarithmic negativity under amplitude damping, thereby establishing the hierarchy
$\mathcal{S}_{A\to B} \;<\; \mathcal{L}_N \;<\; C_{l_1}$
in terms of resistance to decoherence.

A qualitatively different behavior is observed in the phase flip (PF) channel. Quantum steering is suppressed in an intermediate range of $\tau$, vanishing around $\tau\in [0.412,0.588]$, while logarithmic negativity and quantum coherence exhibit symmetric profiles with a minimum at the same value of $\tau$. Notably, in both the PF and PD channels, the overlap between the logarithmic negativity and coherence curves highlights their identical dynamical response to phase-type noise. Overall, these results demonstrate that steering is the most fragile quantum resource, entanglement shows intermediate robustness, and coherence remains the most resilient, with the detailed behavior strongly depending on the nature of the decoherence channel and the experimental configuration.

For the phase damping (PD) channel, quantum steering decreases smoothly with increasing $\tau$ and approaches zero only in the strong decoherence regime, around $\tau \simeq 1$, for all experiments. In this channel, logarithmic negativity and quantum coherence exhibit identical linear decay behaviors, indicating that both quantities are affected in the same manner by the phase damping process. Moreover, KamLAND consistently preserves larger values of these quantities compared to Daya Bay and JUNO.

\section{Dephasing effet and dynamical state}\label{sec:6}

We now describe the dynamical evolution of the two-qubit state when it is subjected to a correlated dephasing quantum channel. The time-dependent density matrix is obtained through the Kraus representation of a completely positive and trace-preserving (CPTP) map, which is given by \cite{NielsenChuang2010,HuFan2020,HuZhou2019}
\begin{equation}
\tilde{	\varrho}_{AB}(t)=\sum_{i,j=0}^{3} L_{ij}\,{	\varrho}_{AB}(0)\,L_{ij}^{\dagger},
	\label{eq:kraus_map}
\end{equation}
where the Kraus operators are defined by
\begin{equation}
	L_{ij}=\sqrt{p_{ij}}\left(\sigma_i\otimes\sigma_j\right),
\end{equation}
with $\sigma_i$ $(i=0,1,2,3)$ denoting the Pauli matrices and $\sigma_0$ the identity operator. The joint probability distribution $p_{ij}$ accounts for classical correlations between the noise channels and is given by \cite{MacchiavelloPalma2002}
\begin{equation}
	p_{ij}=(1-\mu)\,p_i p_j+\mu\,p_i\,\delta_{ij},
	\label{eq:joint_prob}
\end{equation}
where $\mu\in[0,1]$ is the correlation parameter. The limits $\mu=0$ and $\mu=1$ correspond to fully uncorrelated and fully correlated dephasing channels, respectively.
For a pure dephasing process, only the operators $\sigma_0$ and $\sigma_3$ contribute, with probabilities $p_0=1-p$ and $p_3=p$. The time-dependent dephasing probability is governed by a random telegraph noise process \cite{HuZhou2019}, leading to
\begin{equation}
	p(t)=\frac{1-h(t)}{2}.
	\label{eq:pt}
\end{equation}
The decoherence function $h(t)$ takes the form
\begin{equation}
	h(t)=
	\begin{cases}
		e^{-t/(2\chi)}
		\left[
		\cos\!\left(\dfrac{\upsilon t}{2\chi}\right)
		+\dfrac{1}{\upsilon}
		\sin\!\left(\dfrac{\upsilon t}{2\chi}\right)
		\right], & 4\chi>1, \\[2ex]
		e^{-t/(2\chi)}
		\left[
		\cosh\!\left(\dfrac{\upsilon t}{2\chi}\right)
		+\dfrac{1}{\upsilon}
		\sinh\!\left(\dfrac{\upsilon t}{2\chi}\right)
		\right], & 4\chi<1,
	\end{cases}
	\label{eq:ht}
\end{equation}
where
\begin{equation}
	\upsilon=\sqrt{|4\chi^2-1|},
\end{equation}
and $\chi$ denotes the environmental correlation time. As a result of correlated dephasing, the off-diagonal elements of the density matrix are attenuated by the factor
\begin{equation}
	\zeta(t)=(1-\mu)\,h^2(t)+\mu,
	\label{eq:zeta}
\end{equation}
which explicitly shows that increasing classical correlations enhances the robustness of quantum coherence. Consequently, the evolved thermal density matrix preserves its X-shaped structure and can be expressed as
\begin{align}
	\tilde{\varrho}_{AB}(t)=
	\begin{pmatrix}
	0 & 0 & 0 & 0 \\
		0 & \tilde{\varrho}_{22} & \tilde{\varrho}_{23} & 0 \\
		0 & \tilde{\varrho}_{32} & \tilde{\varrho}_{33} & 0 \\
		0 & 0 & 0 & 0
	\end{pmatrix},
\end{align}
where
\begin{align}
	\tilde{\varrho}_{22}&=\varrho_{22},\\\nonumber
	\tilde{\varrho}_{33}&=\varrho_{33},\\\nonumber
	\tilde{\varrho}_{23}&=\zeta(t)\varrho_{23},\\\nonumber
	\tilde{\varrho}_{32}&=\zeta(t)\varrho_{32}.
	\label{eq:rho_t}
\end{align}
This formulation highlights that correlated dephasing selectively suppresses quantum coherences while leaving the population terms invariant. In particular, stronger classical correlations ($\mu\rightarrow 1$) significantly reduce decoherence effects, thereby preserving quantum coherence and quantum correlations over longer time scales.
\begin{figure}[H]
	\includegraphics[scale=0.42]{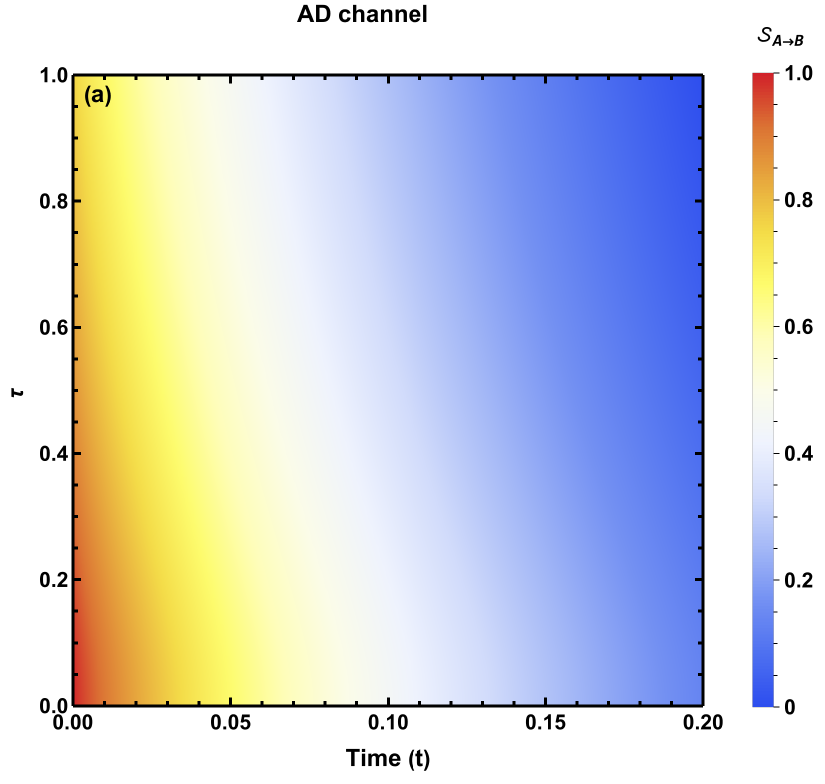}
	\includegraphics[scale=0.42]{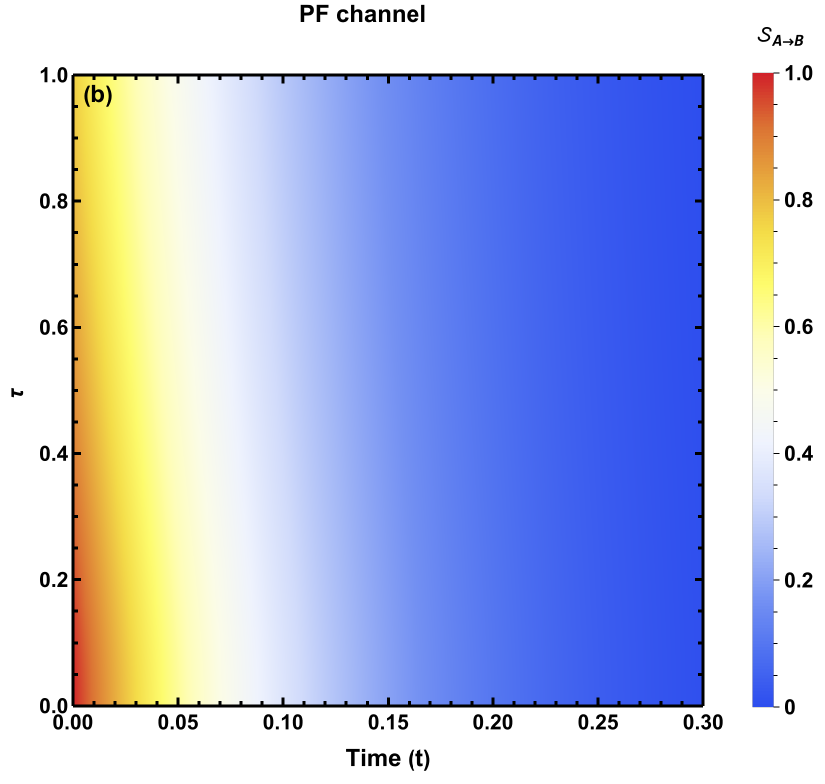}
	\includegraphics[scale=0.42]{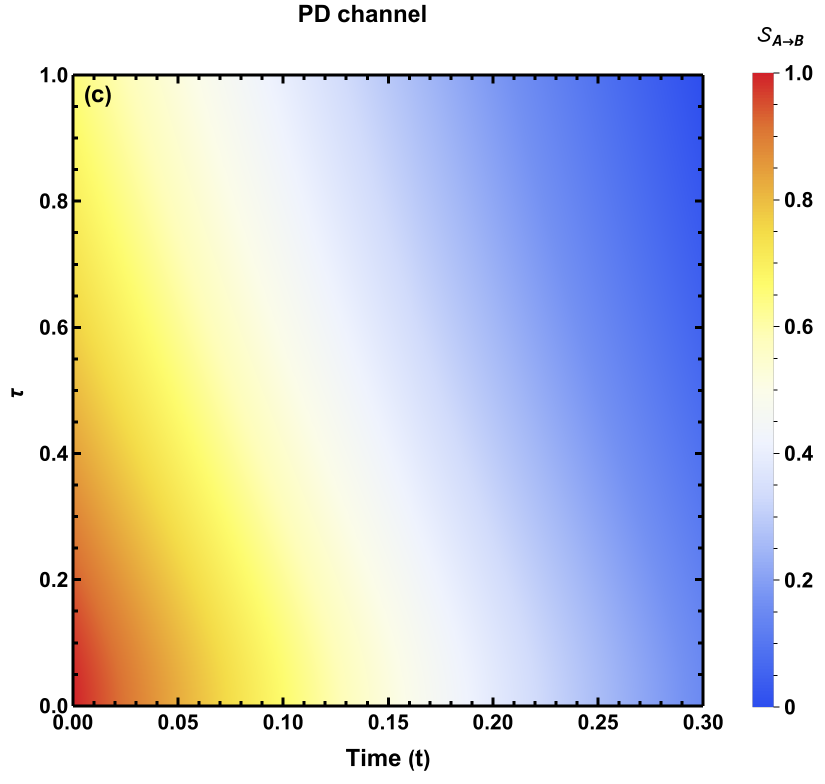}
	\newline
	\vspace*{0.3cm}
	\includegraphics[scale=0.42]{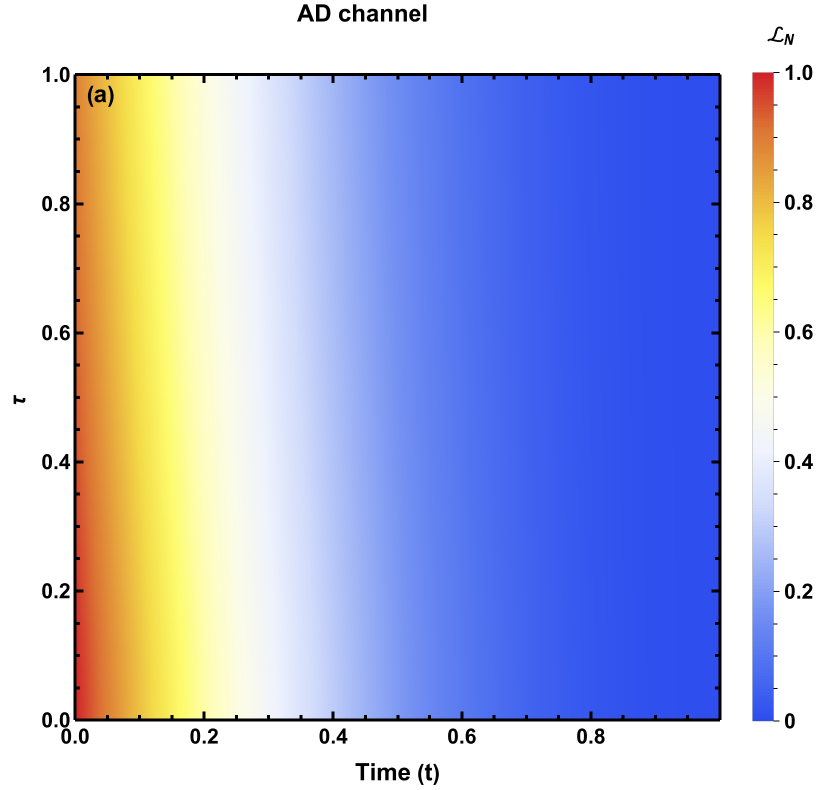}
	\includegraphics[scale=0.42]{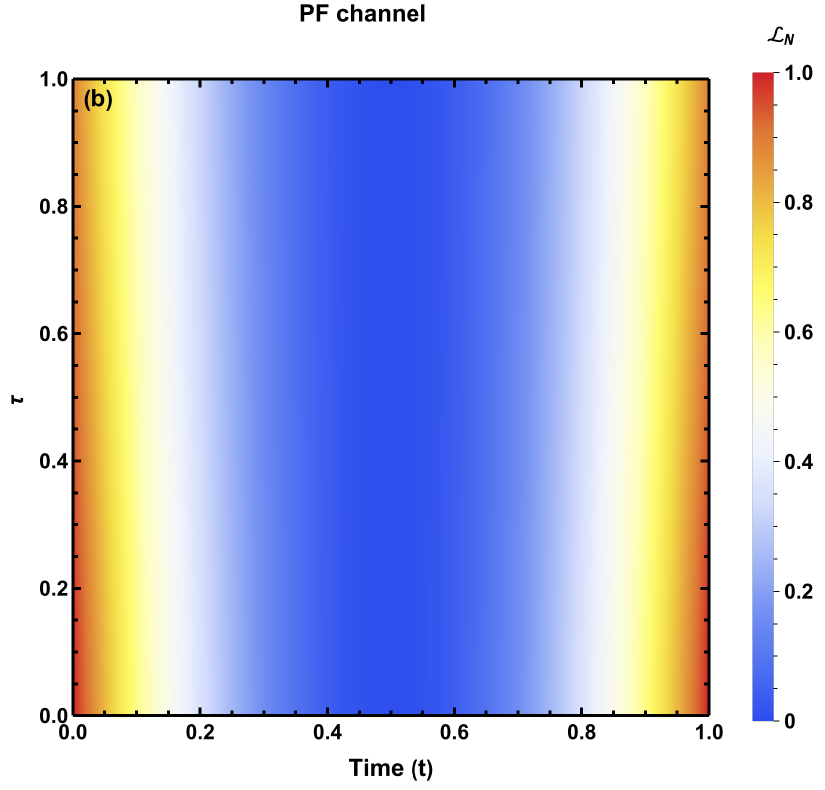}
	\includegraphics[scale=0.42]{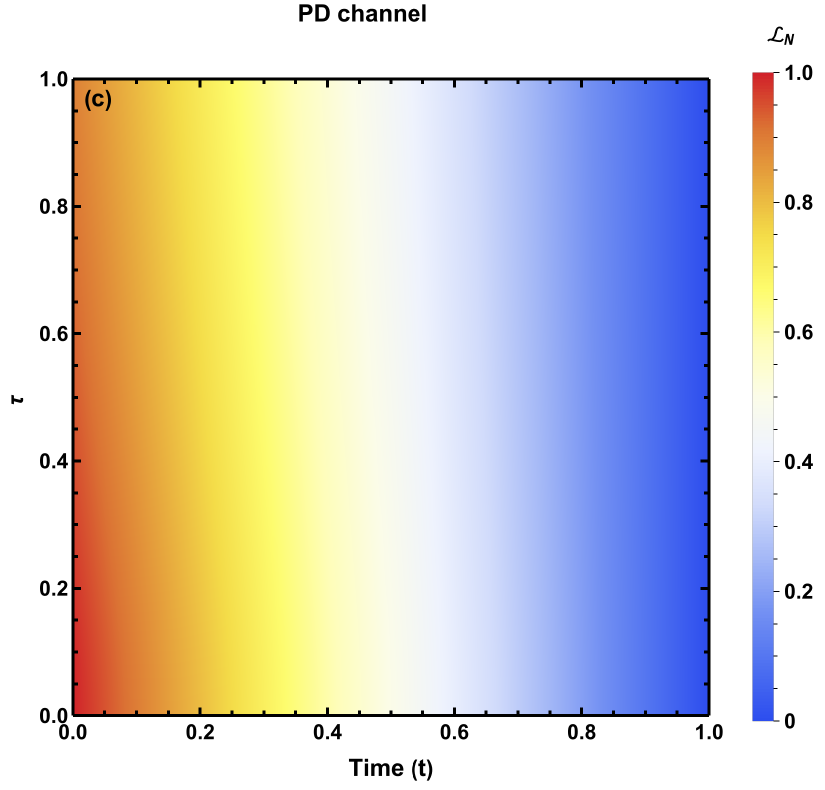}
		\newline
			\vspace*{0.2cm}
	\includegraphics[scale=0.42]{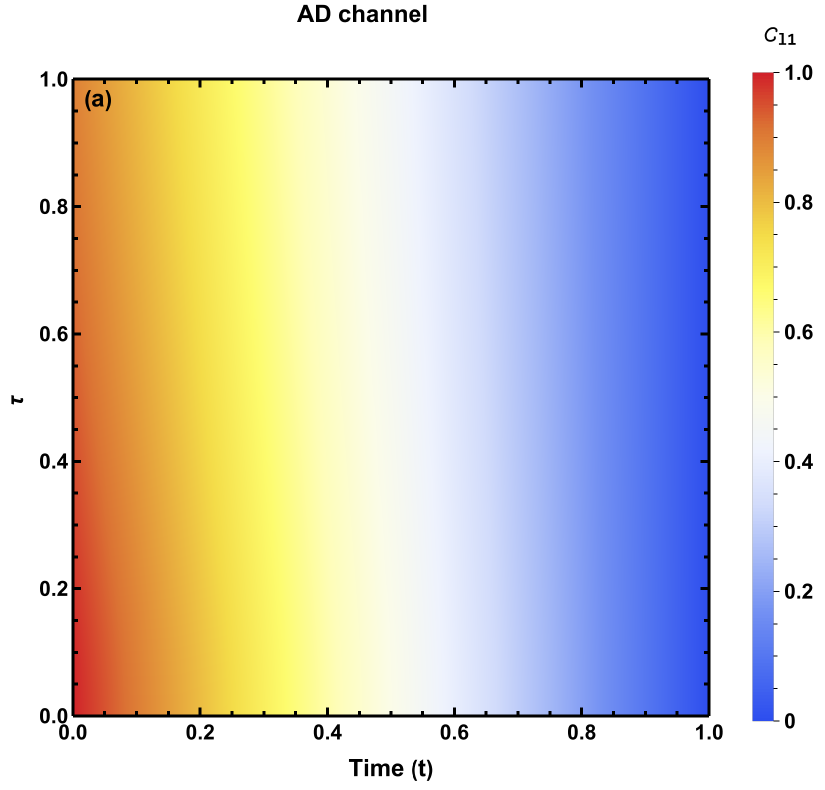}
	\includegraphics[scale=0.42]{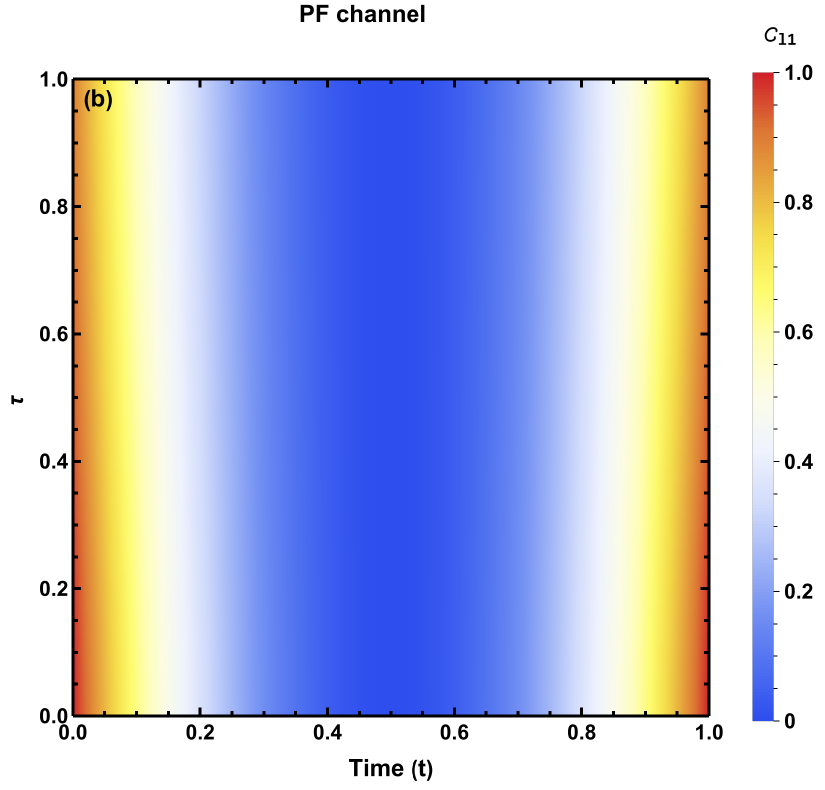}
	\includegraphics[scale=0.42]{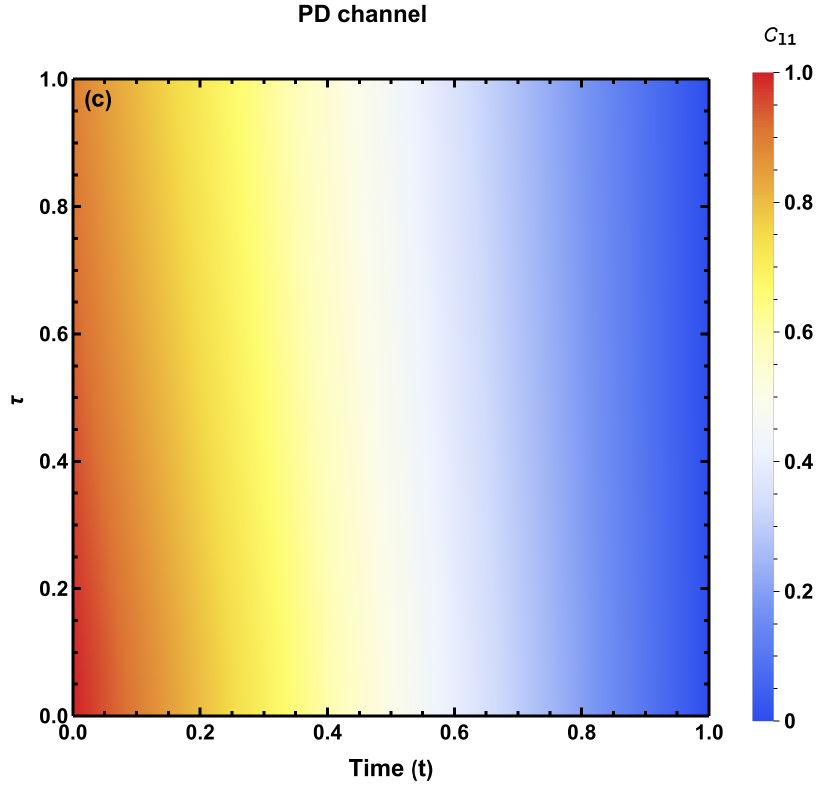}
	\caption{Dynamical evolution of quantum steering, logarithmic negativity, and quantum coherence in the Markovian regime as functions of the decoherence parameter $\tau$ for the KamLAND experiment under the amplitude damping (AD), phase flip (PF), and phase damping (PD) channels, with fixed parameters $\chi = 0.1$ and $\mu = 0.8$.}
	\label{fig:9}
\end{figure}

A global view of Fig.~\ref{fig:9} reveals the joint temporal and decoherence-induced evolution of quantum steering, logarithmic negativity, and quantum coherence for the KamLAND experiment in the Markovian regime. The two-dimensional density maps illustrate how these quantum resources are progressively affected by the interplay between the evolution time $t$ and the decoherence parameter $\tau$ under the AD, PF, and PD channels.

For the amplitude damping (AD) channel, all three quantities exhibit a clear monotonic decay with time, which becomes more pronounced as $\tau$ increases. We remark that the quantum steering is particularly fragile, vanishing rapidly beyond short interaction times even for moderate decoherence strengths, while logarithmic negativity persists over a broader time window. Furthermore, the quantum coherence displays the highest robustness, maintaining nonzero values over longer times and across the full range of $\tau$.

In contrast, the phase flip (PF) channel induces a qualitatively different behavior for entanglement and coherence. Both logarithmic negativity and coherence show symmetric profiles with respect in time, characterized by a minimum around intermediate times and a revival at later stages, reflecting the phase-sensitive nature of the PF noise. Notably, in this channel, the coherence and logarithmic negativity share identical distributions, indicating a direct correspondence between these two resources. Moreover, the quantum steering, however, remains strongly suppressed and vanishes over most of the parameter space.

A similar correspondence between logarithmic negativity and coherence is also observed for the phase damping (PD) channel. Here, both quantities decay smoothly with time and increasing $\tau$, following the same functional behavior, while quantum steering is rapidly destroyed and remains negligible throughout the evolution. These results emphasize the hierarchical robustness of quantum resources, with steering being the most sensitive to decoherence, and highlight the distinct impact of dissipative and dephasing mechanisms on the dynamics of nonclassical correlations.

\begin{figure}[H]
	\includegraphics[scale=0.42]{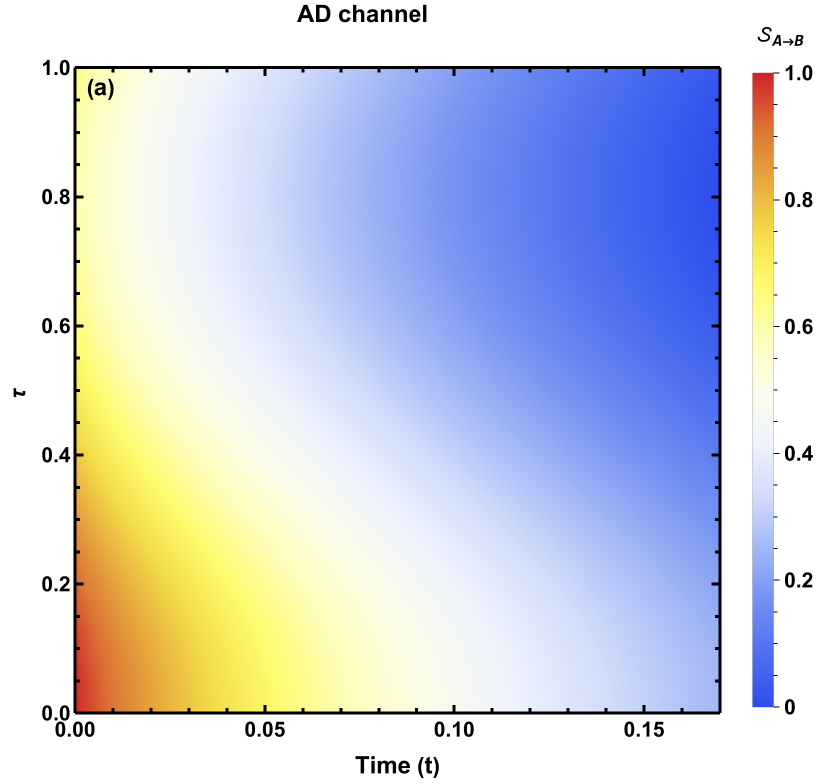}
	\includegraphics[scale=0.42]{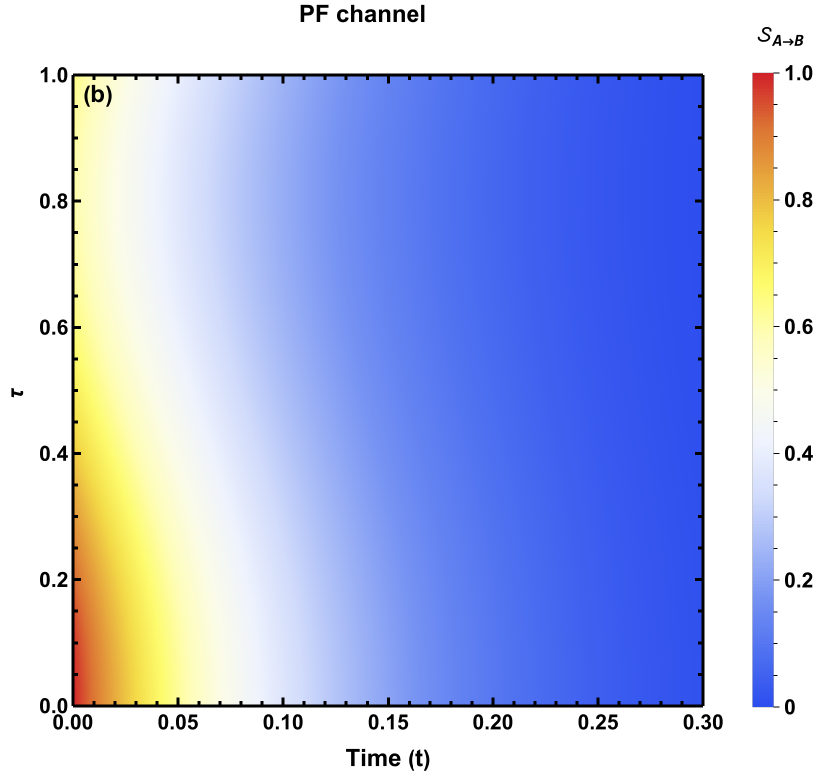}
	\includegraphics[scale=0.42]{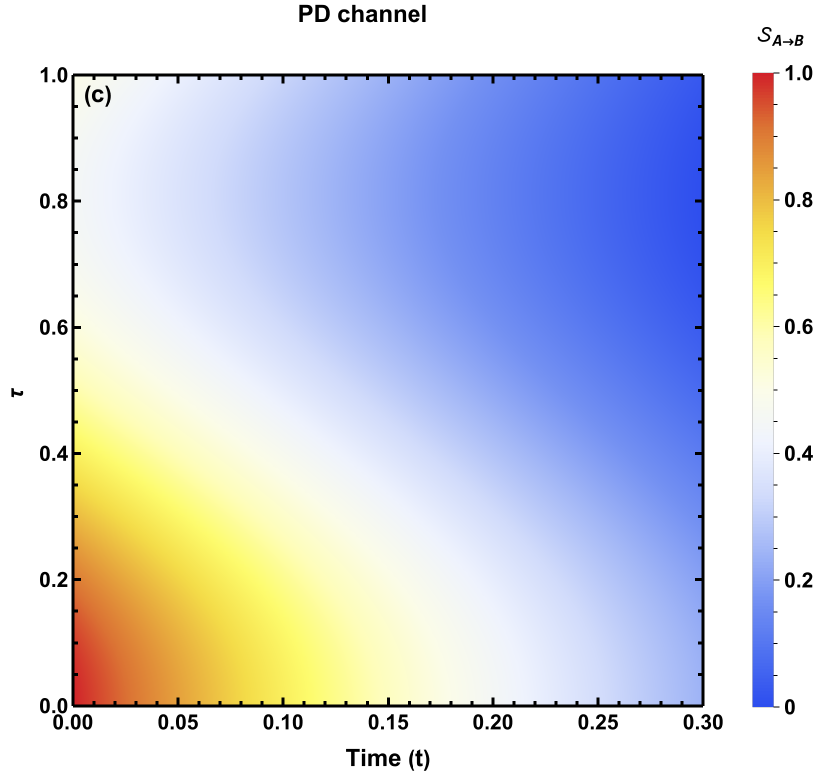}\\
	\includegraphics[scale=0.42]{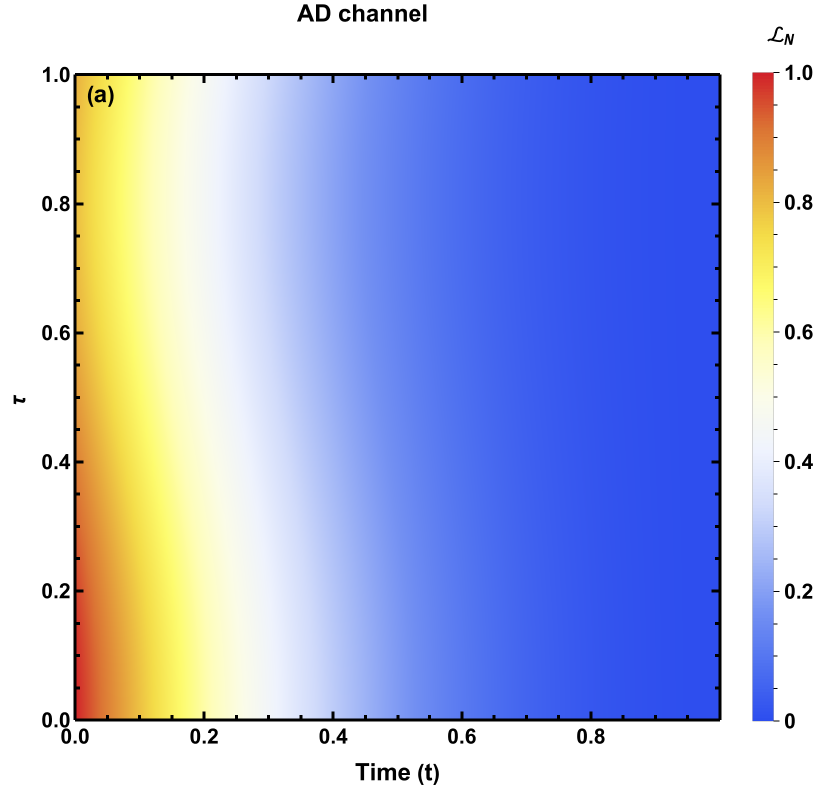}
	\includegraphics[scale=0.42]{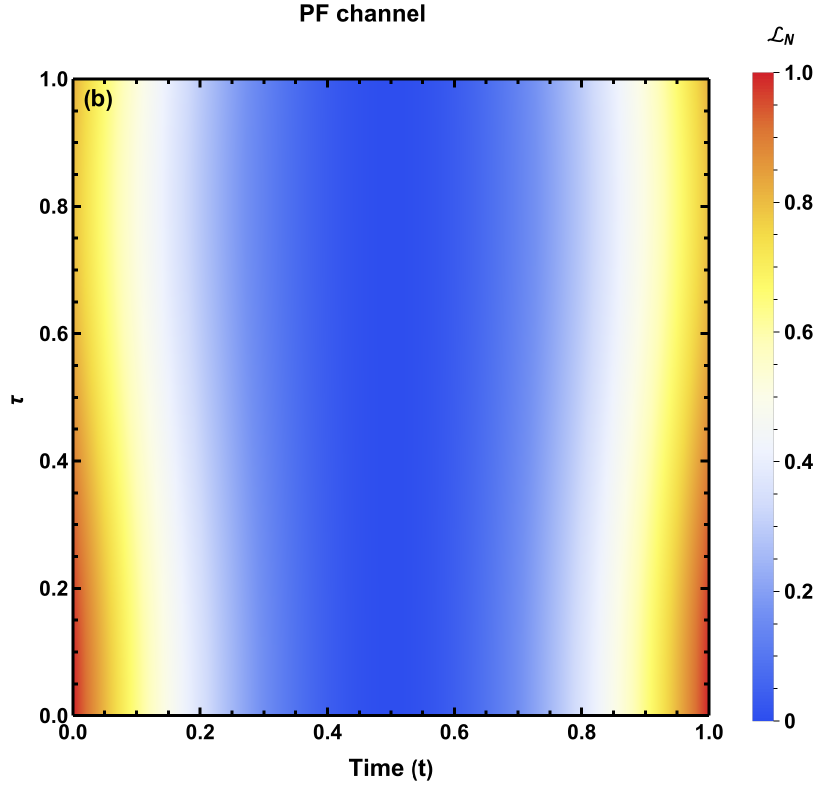}
	\includegraphics[scale=0.42]{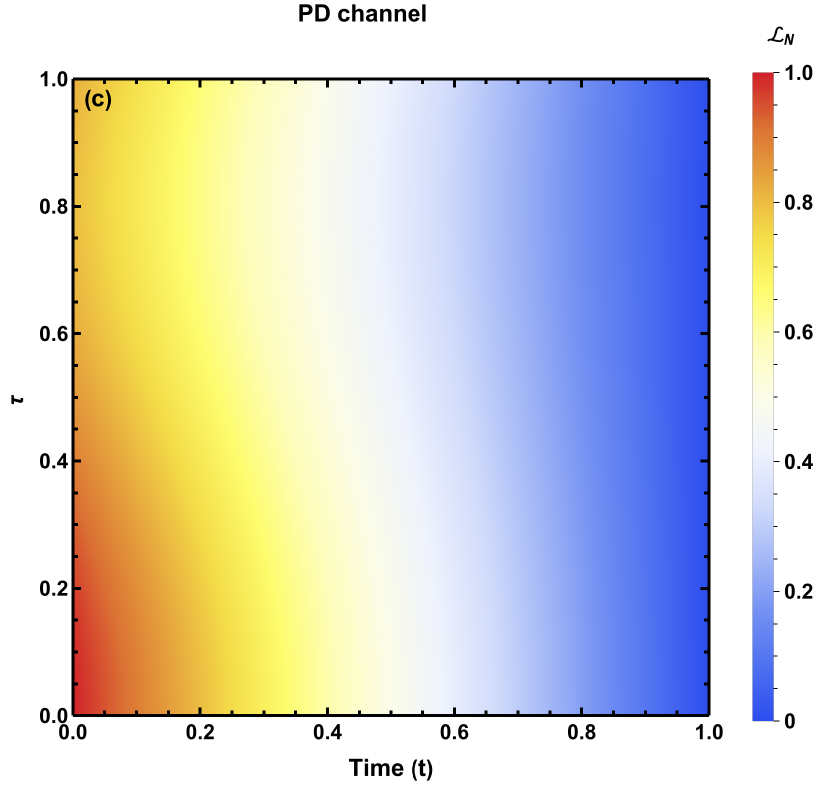}\\
	\includegraphics[scale=0.42]{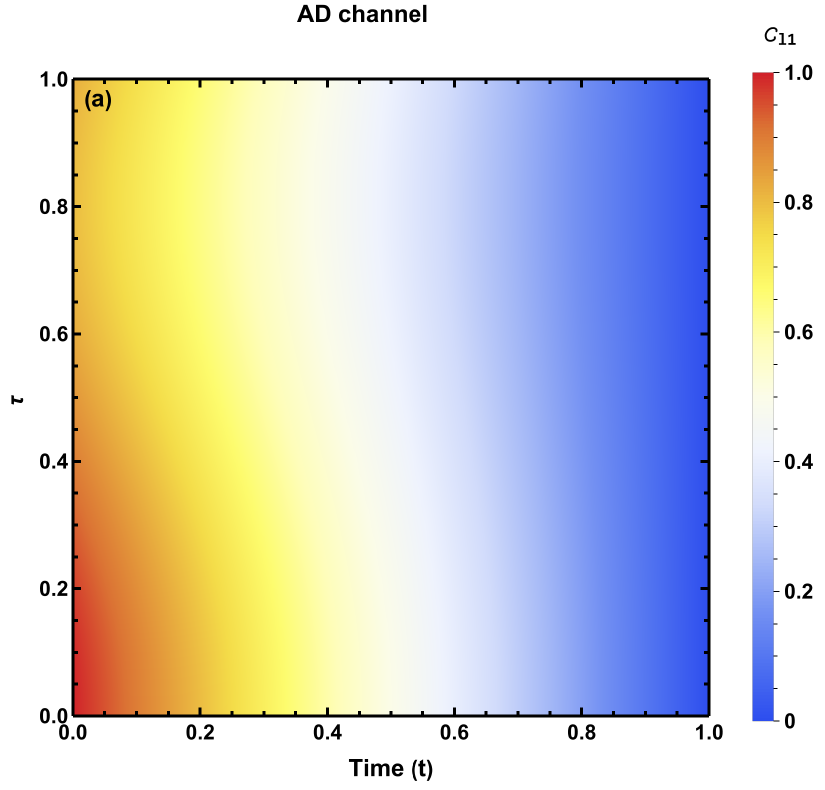}
	\includegraphics[scale=0.42]{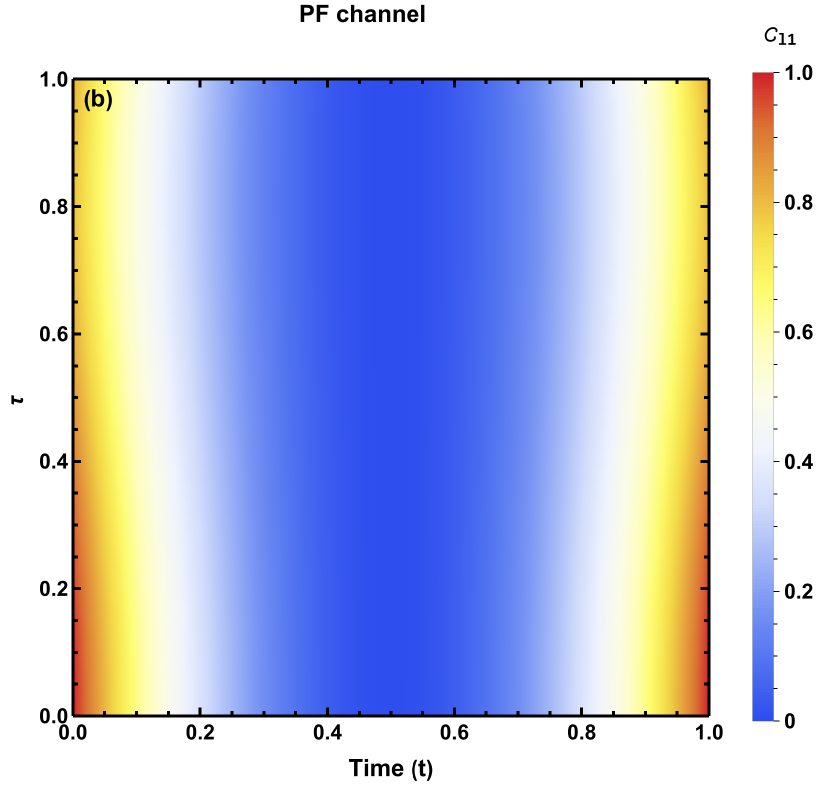}
	\includegraphics[scale=0.42]{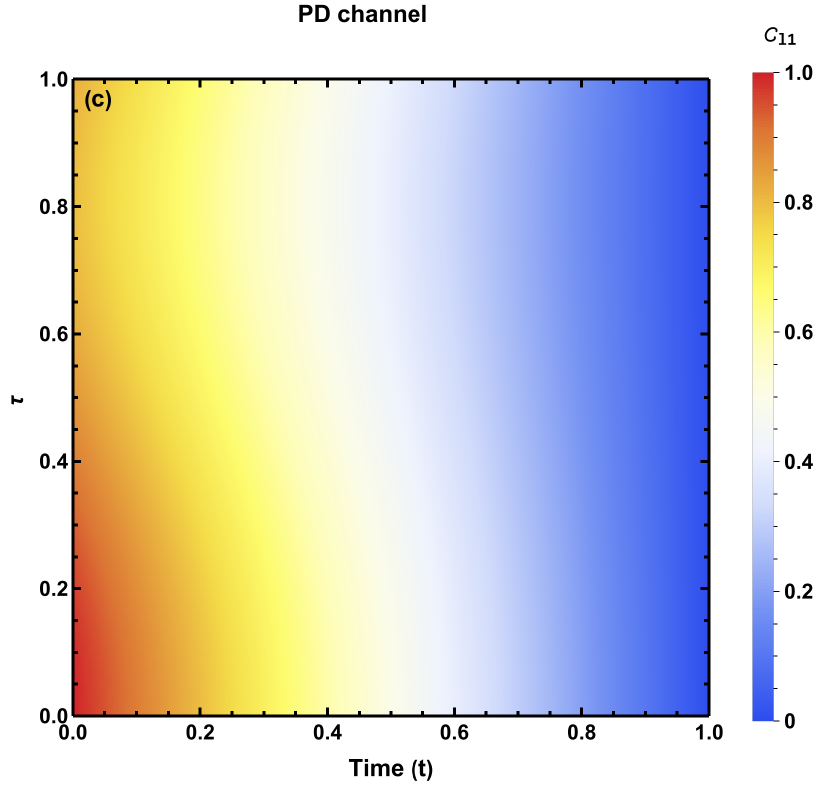}
	\caption{Dynamical evolution of quantum steering, logarithmic negativity, and quantum coherence in the non-Markovian regime as functions of the decoherence parameter $\tau$ for the KamLAND experiment under the amplitude damping (AD), phase flip (PF), and phase damping (PD) channels, with fixed parameters $\chi = 5$ and $\mu = 0.8$.}
	\label{fig:10}
\end{figure}

Figure~\ref{fig:10} illustrates the non-Markovian dynamics of quantum steering, logarithmic negativity, and quantum coherence for the KamLAND experiment as functions of time $t$ and the decoherence parameter $\tau$ under the AD, PF, and PD channels. In contrast to the Markovian case, the presence of memory effects significantly modifies the temporal behavior of all quantum resources.

Under the amplitude damping (AD) channel, the combined influence of non-Markovianity and dissipation leads to a slower degradation of quantum correlations. Quantum steering remains the most sensitive to environmental effects; however, its decay is partially mitigated, and nonzero steerability persists over extended time intervals for small and intermediate values of $\tau$. In addition, logarithmic negativity and quantum coherence exhibit enhanced robustness, highlighting the role of information backflow in compensating dissipative losses.

For the phase flip (PF) channel, pronounced non-Markovian signatures emerge. Both logarithmic negativity and coherence display clear revival patterns in time, characterized by symmetric distributions and temporal oscillations that are absent in the Markovian regime. These revivals indicate a recurrent exchange of information between the system and its environment. In contrast, quantum steering remains largely suppressed, confirming its fragility even in the presence of memory effects.

A similar qualitative behavior is observed for the phase damping (PD) channel. While quantum steering is rapidly diminished and remains negligible across most of the parameter space, logarithmic negativity and coherence experience a slower decay and retain appreciable values over longer times. The close correspondence between entanglement and coherence persists, emphasizing that non-Markovian memory effects primarily enhance these two resources.

Overall, these results demonstrate that non-Markovianity substantially alters the dissipative dynamics by delaying decoherence and enabling partial revivals of quantum resources, with entanglement and coherence benefiting more strongly from memory effects than quantum steering.

\section{Conclusion}   \label{sec:8}

In summary we have investigated the behavior of quantum steering, logarithmic negativity, and quantum coherence in the context of two-flavor neutrino oscillations by modeling the system as an open quantum system. By incorporating realistic environmental effects through amplitude damping (AD), phase flip (PF), and phase damping (PD) channels, we have provided a comprehensive analysis of how different decoherence mechanisms influence the dynamics of quantum resources. Our results revealed a clear hierarchy in the robustness of quantum resources against decoherence. Moreover, the quantum steering emerges as the most fragile quantity, vanishing rapidly as decoherence strength increases in both Markovian and non-Markovian regimes. Besides, we have shown that the logarithmic negativity exhibits intermediate robustness, while quantum coherence to be the most resilient, persisting over a wider range of parameters. In particular, we have shown for the PF and PD channels, the logarithmic negativity and the quantum coherence are identical dynamical behavior, reflecting their common sensitivity to phase-related noise. The comparison between Markovian and non-Markovian dynamics highlighted the crucial role of environmental memory effects. While Markovian evolution led to a monotonic decay of all quantum resources, non-Markovianity significantly delayed decoherence and induced partial revivals of entanglement and coherence due to information backflow. Despite this enhancement, quantum steering remained strongly suppressed, underscoring its vulnerability to environmental disturbances. Overall, this study bridged concepts from quantum information theory and neutrino physics, offering new insights into the persistence and degradation of nonclassical features in neutrino oscillation systems. Our findings contributed to a deeper understanding of quantum correlations in realistic physical settings and provided useful guidance for future investigations of quantum effects in particle physics and other open quantum systems.

\end{document}